\def\draftversion{false}
\RequirePackage{ifthen}
\ifthenelse{\equal{\draftversion}{true}}{
  \documentclass[galley,aps,pra,10pt,amsmath,amssymb,
    superscriptaddress,nofootinbib,longbibliography]{revtex4-2}
}{
  \documentclass[twocolumn,aps,pra,10pt,amsmath,amssymb,
    superscriptaddress,nofootinbib,longbibliography]{revtex4-2}
}
\usepackage{graphicx}% Include figure files
\usepackage[usenames,dvipsnames]{color} % colors
\usepackage{bm} % bold math
\usepackage{soul} % \st    for strike-out
\usepackage{mathtools}
\usepackage{tabularx}
\usepackage{dcolumn}
\usepackage{bbold}
\usepackage{hyperref}
\hypersetup{
    colorlinks=true,       
    linkcolor=blue,          
    citecolor=blue,       
    filecolor=magenta,      
    urlcolor=blue          
}

\ifthenelse{\equal{\draftversion}{true}}{
  \marginparwidth 2.7in
  \marginparsep 0.5in
  \newcounter{comm} 
  \def\commnext{\stepcounter{comm}}
  \def\commtext{{\bf\color{blue}[\arabic{comm}]}}
  \def\commmar{{\bf\color{blue}[\arabic{comm}]}}
  \def\dvm#1{\commnext\marginpar{\small DV\commmar: #1}\commtext}
  \def\tcm#1{\commnext\marginpar{\small TC\commmar: #1}\commtext}
  
  \newcommand{\seclab}[1]{\label{sec:#1}{\Red{\small\;\;[sec:~#1]}}}
  \newcommand{\eqlab}[1]{\Red{\hbox{\small\;\;[#1]}}\label{eq:#1}}
  
}{
  \def\dvm#1{}
  \def\tcm#1{}
  
  \newcommand{\eqlab}[1]{\label{eq:#1}}
  \newcommand{\seclab}[1]{\label{sec:#1}}
  
}

\newcommand{\beq}{\begin{equation}}
\newcommand{\eeq}{\end{equation}}
\newcommand{\bea}{\begin{eqnarray}}
\newcommand{\eea}{\end{eqnarray}}

\newcommand{\eq}[1]{Eq.~(\ref{eq:#1})}

\newcommand{\sref}[1]{Sec.~\ref{sec:#1}}

\newcommand{\ket}[1]{\vert#1\rangle}

\newcommand{\code}[1]{\textsc{#1}}

\begin{document}

\title{Reduced Wannier representation for topological bands}

\author{Trey Cole}
\affiliation{
Department of Physics \& Astronomy, Rutgers University,
Piscataway, New Jersey 08854, USA}

\author{David Vanderbilt}
\affiliation{
Department of Physics \& Astronomy, Rutgers University,
Piscataway, New Jersey 08854, USA}

\begin{abstract}

Bands with non-trivial topological indices have a topological obstruction preventing them from being represented by exponentially localized Wannier states. Here, we propose a procedure to construct exponentially localized Wannier functions that span a subspace of topologically obstructed bands through the use of the projection method. These Wannier functions form what we refer to as a ``reduced Wannier representation," indicating that the Wannier functions necessarily do not span the full topological manifold. By constructing supercells, we obtain reduced Wannier functions that break the primitive translational symmetry while capturing a substantial portion of the topologically obstructed manifold. This approach effectively decomposes the topological manifold into two subspaces: an itinerant subspace inheriting the topology and a localized trivial subspace represented by reduced Wannier functions. We consider the Haldane and Kane-Mele tight-binding models as the platforms for our investigation.

\end{abstract}

\maketitle

\section{Introduction}

Hamiltonians with translational invariance lead to a momentum space description in which the energy eigenstates take the form of Bloch waves labeled by crystal momentum. These states are inherently delocalized throughout the crystal, often obscuring properties that we expect to occur locally in insulators, such as electromagnetic responses and chemical bonding. On the other hand, Wannier functions provide a localized real-space representation of Bloch functions, offering additional insight into the chemical and electronic properties of materials \cite{wannier1937, kohn1959, foster1960, boys1960,cloizeaux1964, strinati1978}. They are frequently used to construct effective tight-binding models, particularly in strongly correlated systems, enabling a simplified yet accurate description of the electronic interactions \cite{solovyev2006, jung2013, gresch2018, lihm2019, Li_2021}. In first-principles calculations, they facilitate the efficient interpolation of $k$-dependent quantities across the Brillouin zone, greatly enhancing computational efficiency \cite{yates2007, calzolari2004}. Moreover, Wannier functions have had a significant role in determining the topology of band structures, aiding in the discovery and characterization of topological materials \cite{kruthoff2017, Bradlyn2017, po2019, Po2018}.

In topological insulators, however, the construction of Wannier functions is inhibited by a ``topological obstruction." When the valence bands possess a nontrivial topological index, such as a Chern number in quantum anomalous Hall insulators or $\mathbb{Z}_2$ invariant in quantum spin Hall insulators, the corresponding Wannier functions fail to be exponentially localized and instead exhibit power law decay \cite{Bradlyn2017, Po2018}. Since Wannier functions are related to Bloch states through a Fourier transform, they require a smooth and periodic gauge to be well localized \cite{strinati1978}. In reciprocal space, the obstruction arises from the inability to choose a gauge for the Bloch eigenstates that simultaneously respects the protecting symmetry and is a smooth and periodic function of wavevector. Specifically, a non-zero Chern number implies that the phase of the Bloch eigenstates winds non-trivially around the Brillouin zone (BZ), leading to vortices or singularities in the phase \cite{thronhauser2006,  soluyanov2011, Vanderbilt2018}. 

From a real-space perspective, the obstruction is rooted in the fact that topological insulators cannot be adiabatically connected to atomic insulators without closing the energy gap. Recent developments have provided a systematic framework for characterizing topological insulators in terms of elementary band representations derived from an atomic limit \cite{kruthoff2017, Bradlyn2017, Cano2018, Cano2021}. In this context, elementary band representations can either form connected bands across the BZ, allowing them to be representable by localized Wannier functions respecting the crystal symmetries, or they can be disconnected, signifying the presence of a topological obstruction \cite{Cano2018}.

Despite these limitations, there have been efforts to construct a Wannier basis in the presence of topological obstructions. Soluyanov and Vanderbilt \cite{soluyanov2011} showed that it is possible to construct localized Wannier functions for $\mathbb{Z}_2$ topological insulators in the topological ($\mathbb{Z}_2$-odd) phase by choosing a gauge that breaks the protecting time-reversal symmetry such that the Wannier functions no longer form Kramers pairs. Gunawardana et al. \cite{Gunawardana2024} recently proposed a method to construct optimally algebraically localized Wannier functions for single-band Chern insulators. Their approach departs from the conventional requirement of exponential localization, instead retaining the vortex in the gauge field, resulting in Wannier functions with logarithmic divergence in the spatial variance. By optimizing the location of the vortices, they minimize the non-divergent part of the spatial variance and construct Wannier functions with optimal power-law decay.

There are systems whose topology is not classified by conventional $K$-theoretic topological indices but still exhibit an obstruction to being Wannier representable. These systems are said to have a ``fragile topology" when the obstruction can be resolved by incorporating additional trivial bands corresponding to an atomic insulator \cite{bouhon2019, po2019, Po2018}. For instance, Song and Bernevig constructed a Wannier representation of two flat bands in magic-angle twisted bilayer graphene, possessing a fragile obstruction, by ``borrowing" irreducible representations from higher-energy trivial bands at high-symmetry $k$-points \cite{song2022}. Similarly, in studies of a kagome lattice model, Hu et al.~\cite{Hu2023} and Chen et al.~\cite{chen2023} obtained a Wannier representation of an obstructed flat band by incorporating representations at high-symmetry momenta from a nearby wide band allowing for an elementary band representation associated with localized orbitals to be formed. 

These previous efforts for Wannierizing topological bands have relied on breaking time-reversal symmetry in $\mathbb{Z}_2$-odd insulators, adding trivial bands to address fragile obstructions, or forgoing exponential localization for bands with non-zero Chern numbers. Here, we propose a unifying framework to obtain truly exponentially localized Wannier functions spanning a subspace of the obstructed bands. We begin by selecting a subspace of the topological band space as the target manifold for Wannierization via subspace selection \cite{souza2001}. This process may involve band folding, i.e., breaking of translational symmetry in the gauge, to enlarge the population of starting bands. We then apply a standard projection method to this subspace to obtain well-localized Wannier functions, whose quadratic spread can be minimized using the methods discussed in the following section. To demonstrate this approach, we use the Haldane model in its Chern insulating phase and the Kane-Mele model in the $\mathbb{Z}_2$-odd phase as prototypical examples.

\section{Maximally Localized Wannier Functions via Projection}
\label{sec:MLWF}

In this section, we review the standard approach for obtaining maximally localized Wannier functions guided by projection from a set of initial trial functions. This material constitutes a synopsis of the relevant material from the review by Marzari et al.~\cite{marzari2012}. The operations described below were implemented in an extension of the \code{PythTB} package \cite{pythtb, cole_zenodo}, which was used to generate the results presented in Sec.~\ref{Reduced_WF} and Sec.~\ref{redwf_Z2}.

\subsection{Wannier Functions}

Wannier functions are related to Bloch functions through a Fourier transform,
\begin{equation}
    |\mathbf{R} n\rangle = \frac{V}{(2\pi)^d} \int_{BZ} d\mathbf{k} e^{-i\mathbf{k}\cdot \mathbf{R}} |\psi_{n\mathbf{k}}\rangle,
\end{equation}
where $\mathbf{R}$ is a real-space lattice vector, $V$ is the volume (or area) of the unit cell, $d$ is the dimensionality of the BZ, and $|\psi_{n\mathbf{k}}\rangle$ are the Bloch states. The Bloch states have a gauge freedom, under which they can be transformed as
\begin{equation}
    |\tilde{\psi}_{n\mathbf{k}}\rangle = e^{i\phi_n(\mathbf{k})} |\psi_{n\mathbf{k}}\rangle
\end{equation}
without affecting the physical description of the system. This gauge freedom leads to a nontrivial modification of the Wannier functions, where a change in gauge results in Wannier functions with altered spreads and shapes.

Instead of a single band index $n$, consider a manifold of $M$ bands that are separated in energy from any other bands, possibly with internal degeneracies. In this case, the notion of gauge freedom can be generalized to a unitary freedom that mixes the states within the manifold,
\begin{equation}
    |\tilde{\psi}_{n\mathbf{k}}\rangle = \sum_{m=1}^{M} U_{mn}^{(\mathbf{k})} |\psi_{m\mathbf{k}} \rangle ,
\end{equation}
where $U_{mn}^{(\mathbf{k})}$ is periodic in $\mathbf{k}$. This unitary freedom is what allows for the construction of well-localized Wannier functions. 

The standard measure of localization of the Wannier functions is given by their quadratic spread,
\begin{equation}
    \Omega = \sum_n \left[ \langle \mathbf{0} n | r^2 | \mathbf{0}n \rangle - \langle \mathbf{0} n | \mathbf{r} | \mathbf{0} n \rangle^2 \right].
\end{equation}
This functional can be decomposed into two separate terms,
\begin{equation}
    \Omega = \Omega_\textrm{I} + \widetilde{\Omega},
\end{equation}
where $\Omega_\textrm{I}$ is gauge-independent, reflecting the intrinsic spread of the chosen subspace, and $\widetilde{\Omega}$ is gauge-dependent. The minimization of the spread is usually decomposed into two stages: first, identifying the optimal subspace of Bloch functions that minimize $\Omega_\textrm{I}$, and then finding the unitary rotation within this subspace that minimizes $\widetilde{\Omega}$. If the goal is to construct Wannier functions that span the entire set of isolated bands, minimizing $\Omega$ corresponds to minimizing only $\widetilde{\Omega}$ using the methods of ``maximal localization" \cite{marzari1997}.  

\subsection{Projection}
\seclab{proj}

One approach to initializing the gauge and subspace selection is to project the bands onto a set of localized trial orbitals $g_n(\mathbf{r})$. These trial functions are typically chosen based on chemical intuition for how the manifold's localized states may behave. The resulting Wannier functions will tend to have the character of these trial orbitals and serve as a starting point for further minimization of the quadratic spread. 

Suppose an isolated manifold of $M$ Bloch states, called the ``target manifold," is to be Wannerized. The projection is carried out by projecting the target manifold onto the trial orbitals,
\begin{equation}
    |\Phi_{n \mathbf{k}} \rangle = \sum_{m=1}^M |\psi_{m\mathbf{k}} \rangle \langle \psi_{m\mathbf{k}} | g_n \rangle .
\eqlab{Phink}
\end{equation}
In the summation, the Bloch states appear in both bra and ket forms so that any random phases they carry get canceled out, and the projected states $|\Phi_{n \mathbf{k}} \rangle$ have a smooth gauge in $\mathbf{k}$. From these projected states, one can construct a matrix of inner products, 
\begin{equation}
(A_{\mathbf{k}})_{mn} = \langle \psi_{m\mathbf{k}} | g_n \rangle,
\eqlab{Adef}
\end{equation}
and define an overlap matrix,
\begin{equation}
(S_{\mathbf{k}})_{mn} = \langle \Phi_{m\mathbf{k}} | \Phi_{n\mathbf{k}} \rangle = (A_{\mathbf{k}}^{\dagger} A_{\mathbf{k}})_{mn},
\end{equation}
that can be used to construct a set of Löwdin-ortho\-nor\-malized Bloch-like states that maintain a smooth gauge in $\mathbf{k}$,
\begin{equation}
    \begin{split}
    |\tilde{\psi}_{n\mathbf{k}} \rangle &= \sum_{m=1}^M |\Phi_{m\mathbf{k}} \rangle (S_{\mathbf{k}}^{-1/2})_{mn} \\&= \sum_{m=1}^M |\psi_{m\mathbf{k}} \rangle (A_{\mathbf{k}}S_{\mathbf{k}}^{-1/2})_{mn} .
    \end{split}
    \eqlab{proj}
\end{equation}
These states are ``Bloch-like" in that they transform like Bloch states under translation but are no longer eigenstates of the Hamiltonian.

When choosing a number of trial wave functions equal to the dimension of the target manifold, the Bloch-like states are related to the original Bloch energy eigenstates by a unitary transformation. This can be seen with the singular value decomposition (SVD) of the square matrix of inner products, 
\begin{equation}
\eqlab{SVD}
    A_{\mathbf{k}} = V_{\mathbf{k}} \Sigma_{\mathbf{k}} W_{\mathbf{k}}^{\dagger} ,
\end{equation} 
where $\Sigma = \text{diag}(\sigma_1, \sigma_2, \cdots, \sigma_M)$ is a diagonal matrix of nonnegative singular values and $V$ and $W$ are unitary $M\times M$ matrices. From this decomposition, \eq{proj} can be recast as,
\begin{equation}
|\tilde{\psi}_{n\mathbf{k}} \rangle = \sum_{m=1}^M |\psi_{m\mathbf{k}} \rangle (V_{\mathbf{k}}W_{\mathbf{k}}^{\dagger})_{mn} \;,\;n=1,\cdots,M.
\eqlab{VWBlochlike}
\end{equation}
It is now apparent that this transformation is unitary when $A$ is square and non-singular, and the Bloch-like states are expected to retain the physical properties of the target manifold. If $A$ is singular, then $VW^{\dagger}$ becomes semi-unitary.

If the number of trial wave functions, $J$, is smaller than the dimension of the target manifold, $M$, the inner product matrix $A$ becomes an $M \times J$ rectangular matrix. In this case, $\Sigma$ is an $M\times J$ rectangular diagonal matrix, while $V$ and $W$ are $M\times M$ and $J \times J$ unitary matrices, respectively. Consequently, the matrix $VW^{\dagger}$ appearing in \eq{VWBlochlike} becomes an $M \times J$ semi-unitary matrix. Explicitly, it takes the form $V \mathbb{1}_{M \times J} W^{\dagger}$, projecting the target manifold onto a lower $J$-dimensional subspace as
\begin{equation}
|\tilde{\psi}_{n\mathbf{k}} \rangle = \sum_{m=1}^M |\psi_{m\mathbf{k}} \rangle (V_{\mathbf{k}}\mathbb{1}_{M \times J}W_{\mathbf{k}}^{\dagger})_{mn} \;,\;n=1,\cdots, J .
\eqlab{VWJBlochlike}
\end{equation}

\subsection{Subspace Selection}
\seclab{selection}

In situations where the target manifold is not isolated from unwanted bands, or where one is interested in extracting an optimally smooth subspace from it, a procedure known as subspace selection, or disentanglement, can be used \cite{souza2001}. This allows for the extraction of a $J$-dimensional subspace of a generally momentum-dependent $M_{\mathbf{k}}$-dimensional manifold containing the states to be Wannierized. The chosen subspace is the one that minimizes the gauge-invariant spread $\Omega_\textrm{I}$. In the present context, we optionally use this technique to improve the smoothness and reduce $\Omega_\textrm{I}$ for the $J$-dimensional subspace initially spanned by the Bloch-like states of \eq{VWJBlochlike}.

On a discrete mesh of $N_k$ $k$-points, the gauge-invariant spread can be written as
\begin{equation}
    \Omega_\textrm{I} = \frac{1}{N_k} \sum_{\mathbf{k}, \mathbf{b}} w_b \text{Tr} \left[  P_{\mathbf{k}} Q_{\mathbf{k+b}}  \right] 
\eqlab{OmegaI}
\end{equation}
where $\mathbf{b}$ is a vector connecting neighboring $k$-points, $w_b$ is the weight associated with a finite-difference shell \cite{marzari1997}, and $P_{\mathbf{k}} = \sum_{n=1}^{J} |\tilde{u}_{n\mathbf{k}} \rangle \langle \tilde{u}_{n\mathbf{k}} |$ is the projector onto the selected subspace. The $\tilde{u}_{n\mathbf{k}}$ are initialized as the cell-periodic versions of the Bloch-like states of \eq{VWJBlochlike}, and $Q_{\mathbf{k}} = \mathbb{1} - P_{\mathbf{k}}$ is the complement of $P_{\mathbf{k}}$. From this relationship, it is clear that $\Omega_\textrm{I}$ is a measurement of the degree of subspace mismatch between neighboring $k$-points.\footnote{This is evident from the relation $\text{Tr} \left[  P_{1} Q_{2}  \right] = || P_1 - P_2||^2/2$.} To minimize this part of the spread, the states at each $\mathbf{k}$ must be chosen to be as close in character as possible to those at neighboring $\mathbf{k}$.

Typically, it is beneficial to obtain the initial subspace spanned by the $J$ Bloch-like states from projection. This subspace is iteratively updated to move towards the minimum of $\Omega_\textrm{I}$. Minimizing $\Omega_\textrm{I}$ is equivalent to solving the eigenvalue equation
\begin{equation}
    \left[ \sum_{\mathbf{b}} w_b P_{\mathbf{k+b}}^{(i-1)}\right] |\tilde{u}_{n\mathbf{k}}^{(i)}\rangle = \lambda_{n\mathbf{k}}^{(i)}  |\tilde{u}_{n\mathbf{k}}^{(i)}\rangle
\end{equation}
for the $i$'th iteration and selecting the $J$ eigenvectors with the largest eigenvalues \cite{souza2001, marzari2012}. At this point, it is appropriate to realign the final $J$ Bloch-like functions $|\tilde{u}_{n\mathbf{k}}\rangle$ to the $J$ trial states following the procedure of \sref{proj}. That is, returning to \eq{Phink}, we use the current $\ket{\tilde{\psi}_{m\mathbf{k}}}$ in place of the $\ket{\psi_{m\mathbf{k}}}$ on the right-hand side, replace $M$ by $J$, and repeat the procedure terminating in \eq{VWBlochlike}. These eigenvectors then form the $J$-dimensional manifold from which Wannier functions can be obtained.

\subsection{Maximal Localization}

As a final step, we may find a gauge that minimizes the gauge-dependent part of the spread $\widetilde{\Omega}$. In practice, all one needs is the set of overlap matrices of the Bloch orbitals at neighboring $k$-points,
\begin{equation}
    M_{mn}^{ (\mathbf{k}, \mathbf{b}) } = \langle u_{m \mathbf{k}} | u_{n, \mathbf{k}+\mathbf{b}} \rangle ,
\end{equation}
which encodes the information about first-order derivatives on a discrete $k$-mesh. The first-order change of the spread from an infinitesimal gauge transformation $U_{mn}^{(\mathbf{k})} = \delta_{mn} + dW_{mn}^{(\mathbf{k})}$, where $dW$ is an anti-Hermitian matrix, can be written as \cite{marzari1997, marzari2012}
\begin{equation}
    G^{(\mathbf{k})} \equiv \frac{d\Omega}{dW^{(\mathbf{k})}} = 4 \sum_{\mathbf{b}} w_b \left( \mathcal{A}\left[ R^{(\mathbf{k},\mathbf{b})} \right ] - \mathcal{S}\left[ T^{(\mathbf{k},\mathbf{b})}\right] \right)
    \eqlab{EQ:grad}
\end{equation}
where $\mathcal{A}\left[ \mathcal{O} \right ] = (\mathcal{O} - \mathcal{O}^{\dagger})/2$ and $\mathcal{S}\left[ \mathcal{O} \right ] = (\mathcal{O} + \mathcal{O}^{\dagger})/2i$ are the anti-symmetric and symmetric superoperators respectively, and the quantities in the sum are defined as
\begin{align}
    R^{(\mathbf{k},\mathbf{b})}_{mn} &= M^{(\mathbf{k},\mathbf{b})}_{mn}M^{(\mathbf{k},\mathbf{b}) *}_{nn} , \\
    \begin{split}
    T^{(\mathbf{k},\mathbf{b})}_{mn} &= \frac{M^{(\mathbf{k},\mathbf{b})}_{mn}}{M^{(\mathbf{k},\mathbf{b})}_{nn}} 
    \;\Big( \text{Im} \ln M^{(\mathbf{k},\mathbf{b})}_{nn} \\ & \hspace{1.0cm} - \mathbf{b} \cdot \frac{1}{N_k} \sum_{\mathbf{k}^{\prime}, \mathbf{b}^{\prime}} w_{b^{\prime}} \mathbf{b}^{\prime} \text{Im} \ln M^{(\mathbf{k}^{\prime},\mathbf{b}^{\prime} )}_{nn} \Big) .
    \end{split}
\end{align}
\tcm{Fixed $w_b$ to have prime}
The gradient obtained from \eq{EQ:grad} can be used in a steepest-decent algorithm to find the global minimum. At each iteration, the step
\begin{equation}
    dW^{(\mathbf{k})} = \epsilon G^{(\mathbf{k})}
\end{equation}
is used, where $\epsilon$ is a small positive constant, and the unitary transformation on the Bloch manifold, initialized as the identity, is updated as
\begin{equation}
    U^{(\mathbf{k})} \rightarrow U^{(\mathbf{k})} \exp\left[ dW^{(\mathbf{k})} \right]
\end{equation}
along with the nearest-neighbor overlap matrices 
\begin{equation}
    M^{(\mathbf{k}, \mathbf{b})} \rightarrow U^{(\mathbf{k})\dagger} M^{(\mathbf{k}, \mathbf{b})} U^{(\mathbf{k} + \mathbf{b})} .
\end{equation}
The minimization is complete when the change in spread after subsequent iterations is below a small threshold. Once the minimum has been reached, the unitary transformation can be applied to the Bloch-like states to achieve an optimally smooth gauge.

\subsection{Topological Obstruction}

As discussed earlier, a Chern insulator has a topological obstruction preventing the construction of exponentially localized Wannier functions. In the context of the projection method, the topological obstruction manifests as a rank deficiency of the inner product matrix $A_{\mathbf{k}}$ defined in \eq{SVD}. As the system transitions from a trivial to a topological phase, a band inversion occurs, swapping the character of the highest occupied and lowest unoccupied bands. When attempting to Wannierize the $n_{\rm occ}$-dimensional occupied manifold using a set of $n_{\rm occ}$ trial wave functions, the inner product matrix becomes rank deficient at certain points in the BZ due to this change in band character. This rank deficiency causes one or more singular values to drop to zero, making the normalization factor $S_{\mathbf{k}}^{-1/2} = W_{\mathbf{k}} \Sigma_{\mathbf{k}}^{-1} W_{\mathbf{k}}^{\dagger}$ in \eq{proj} diverge. Consequently, the projection fails to produce well-localized Wannier functions. The approach outlined below aims to address this issue.

\section{Reduced Wannier Functions for Chern Insulators}
\label{Reduced_WF}
\seclab{avoid-topo}

To avoid the obstruction, we abandon the goal of generating a set of Wannier functions that fully spans the topological manifold and instead aim to construct a set that spans ``most of'' the topological manifold. To do so, we reduce the number of trial functions such that the overlap matrix has full rank throughout the BZ, resulting in a set of Bloch-like states spanning a subspace of the topological manifold. This modification enables us to obtain exponentially-localized Wannier functions that can be further optimally localized using the methods of subspace selection and maximal localization.

\subsection{Haldane Model}

The Haldane model is a prototypical example of a Chern insulator and serves as the setting for testing our approach. The model is comprised of a two-dimensional honeycomb lattice, where, after setting the lattice constant $a=1$, the lattice vectors can be chosen as $\mathbf{a}_1 = (1, 0)$ and $\mathbf{a}_2 = (1/2, \sqrt{3}/2)$. The primitive cell contains two spinless orbitals at positions $\mathbf{\tau}_A = (1/3) \mathbf{a}_1 + (1/3) \mathbf{a}_2$ and $\mathbf{\tau}_B = (2/3) \mathbf{a}_1 + (2/3) \mathbf{a}_2$, located on sublattices $A$ and $B$ respectively. The onsite energy is $-\Delta$ for orbitals on sublattice $A$, and $\Delta$ for orbitals on sublattice $B$, defining the low- and high-energy sublattices respectively. The tight-binding parameters consist of real nearest-neighbor hopping $t_1$ and complex next-nearest neighbor hopping $\pm it_2$. In the second quantized formalism, the tight-binding Hamiltonian is
\begin{equation}
    \begin{split}
     H = &\ \Delta \sum_i (-)^{i} c_i^{\dagger} c_i -t_1 \sum_{\langle i, j \rangle} (c_i^{\dagger}c_j + h.c.) \\ &+ t_2 \sum_{\langle\langle i, j \rangle\rangle} (i c_i^{\dagger}c_j +h.c. ) .
    \end{split}
\end{equation}

A gap closure occurs at the $\mathbf{K}$ and $\mathbf{K}^\prime$ points for $t_2 = \pm \Delta / 3\sqrt{3}$ respectively. Henceforth, we set $\Delta = t_1 = 1$, in which case the topological transition occurs when the next-nearest neighbor hopping $t_2$ passes through $|t_2|=1/3\sqrt{3}\approx 0.19$. We present results for the trivial phase at $t_2=-0.1$ and the topological phase with Chern number $C=1$ at $t_2=-0.3$.

\begin{figure}[t!]
\begin{center}
\includegraphics[width=3.4in]{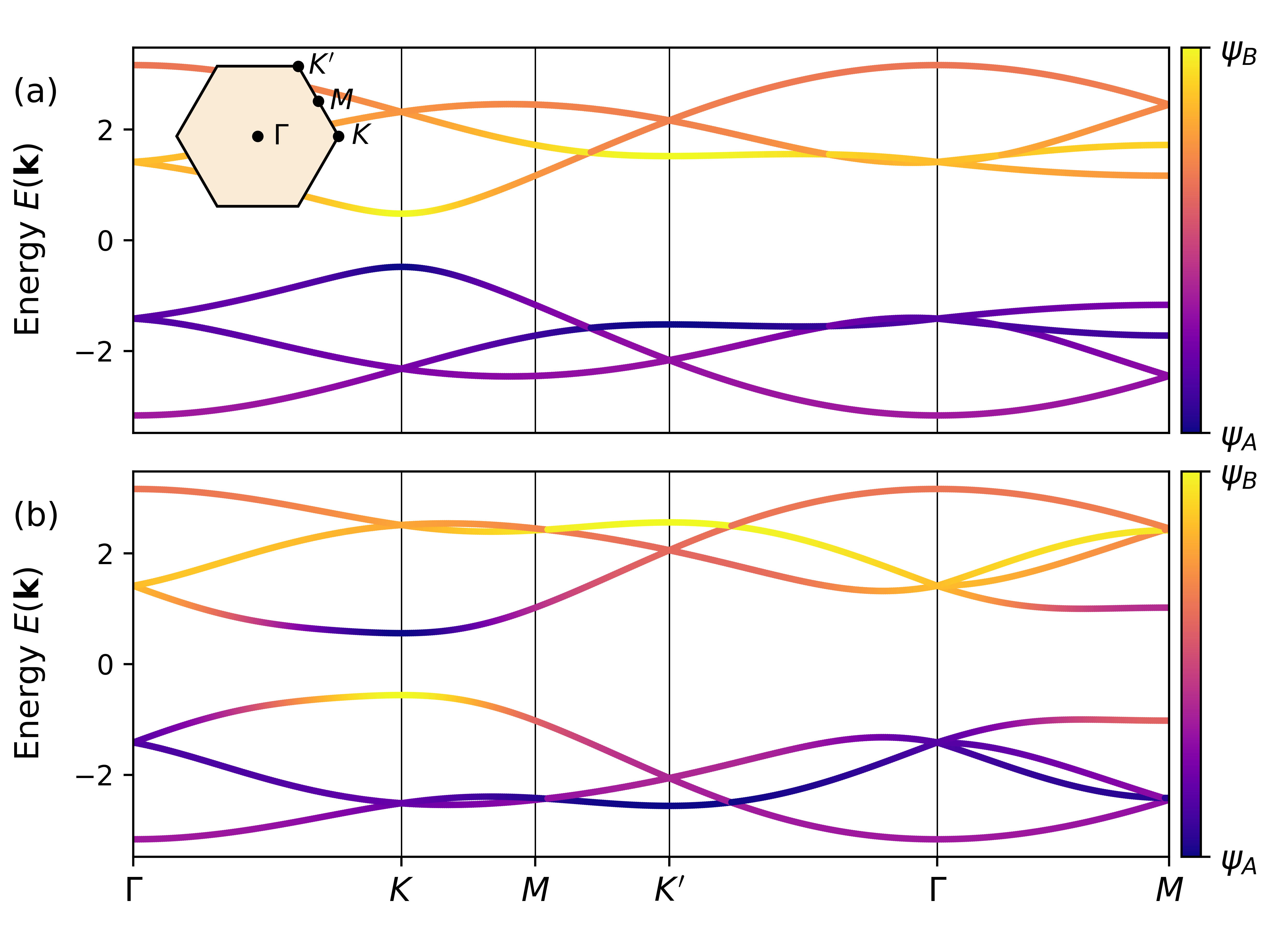}
\end{center}
% \centering
\vspace{-10mm}
\caption{Bands for the Haldane model on a $2\times 2$ supercell. The color indicates the projection of the eigenstates onto sublattices, with purple and yellow representing the low- and high-energy sublattices, respectively. (a) The bands in the trivial phase ($C=0$) where $t_2 = -0.1$. (b) The bands in the topological phase ($C=1$) where $t_2 = -0.3$. In both cases, $\Delta = t_1 = 1$.}
\label{FIG1}
% \vspace{-1mm}
\end{figure}

\subsection{Reduced Wannier Functions}
\label{ss:dfcntWF}

 The Haldane model at half-filling has only one occupied band, which normally corresponds to one Wannier function per primitive cell. However, in the topological phase, we know it is impossible to construct a full set of Wannier functions of this density. As an example of our strategy, we consider the possibility of constructing three Wannier functions for every four primitive cells, thus representing 3/4 of the occupied subspace. That is, we start from a 2$\times$2 supercell, for which the construction of four Wannier functions would normally be straightforward, and instead attempt the construction of three Wannier functions in the topological phase.

Fig.~\ref{FIG1} shows the band structure of the Haldane model computed using the \code{PythTB} package~\cite{pythtb} with a $2\times 2$ supercell on a $20\times 20$ regular $k$-mesh. A band inversion occurs in going from the trivial phase of Fig.~\ref{FIG1}(a) to the topological phase of Fig.~\ref{FIG1}(b) at the $\mathbf{K}$ point, which has been mapped from the $\mathbf{K}^{\prime}$ point after band folding. The color gradient indicates that the characters of the highest occupied and lowest unoccupied bands swap at $\mathbf{K}$; the eigenstate of the highest energy valence band becomes fully localized on the high-energy sublattice, while the eigenstate of the lowest energy conduction band becomes fully localized on the low-energy sublattice. This inversion is a hallmark of the topological phase and signals the nontrivial nature of the band structure.

\begin{figure}[t!]
\begin{center}
\includegraphics[width=3.4in]{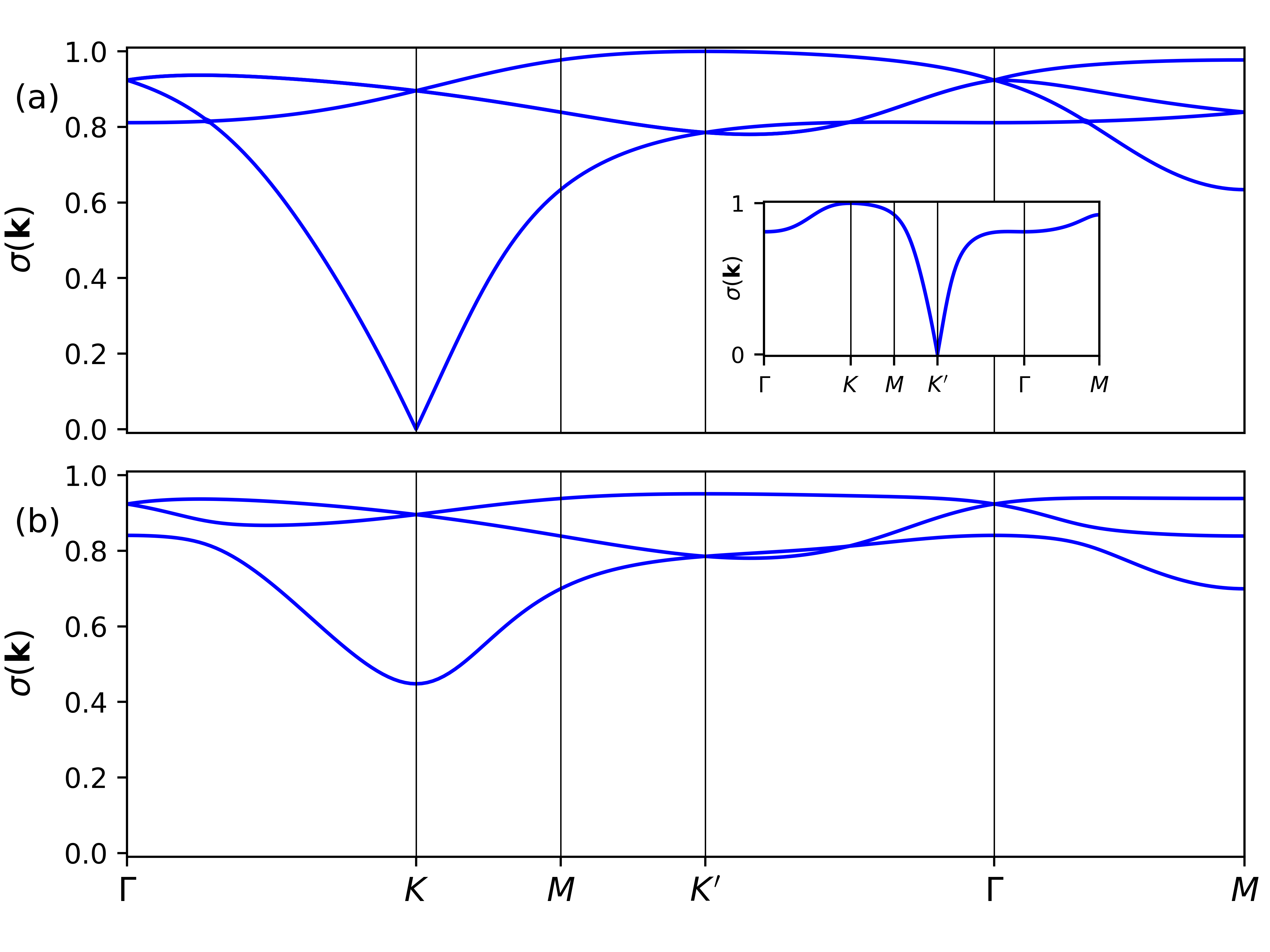}
\end{center}
% \centering
\vspace{-10mm}
\caption{The singular values plotted along a high-symmetry path in the BZ corresponding to the Haldane model in the $C=1$ phase for a $2\times2$ supercell. (a) Four delta functions on the low-energy sites were used as trial wave functions. The inset shows the singular value in the case of a single primitive cell, with the trial wave function being a delta function on the sole low-energy site. (b) Three delta functions on a subset of low-energy sites are used as trial wave functions. }
\label{FIG2}
% \vspace{-1mm}
\end{figure}

Consider that we attempt to apply the projection procedure to Wannierize the complete set of four topologically obstructed valence bands shown in Fig.~\ref{FIG1}(b). We use the natural choice of trial wave functions localized on the four low-energy A sites in the 2$\times$2 supercell, represented by delta functions $g_n(\mathbf{r}) = \delta(\mathbf{r} - \boldsymbol{\tau}_{A,n})$. This yields a square 4$\times$4 inner product matrix $A_{\mathbf{k}}$ that becomes singular at the $\mathbf{K}$ point, where it has an entire column of zeros, 
\begin{equation}
    (A_{\mathbf{K}})_{4,n} = \langle \psi_{4, \mathbf{K}} | g_{n} \rangle = 0
\end{equation}
for all $n$. This results from the fact that the highest-energy occupied state $|\psi_{4, \mathbf{K}}\rangle$ has no support on the low-energy sublattice after the band inversion. Consequently, the inner-product matrix at this point is rank-deficient, with $\text{rank}(A_{\mathbf{K}})=3$, and has one singular value equal to zero, as shown in Fig.~\ref{FIG2}(a). As mentioned, this leads to a divergence in the normalization factor in the unitary projection, shown in \eq{proj}. This rank deficiency reflects the presence of a topological obstruction preventing the construction of localized Wannier functions for the full set of valence bands. Even if we were to choose a different set of trial wave functions, altering the gauge of the Bloch-like states, the singularity would persist and shift to another point in the BZ.  

To resolve the singularity, we reduce the number of trial wave functions so as to make $A_{\mathbf{k}}$ a rectangular matrix with full rank at the problematic $\mathbf{K}$ point.\footnote{A matrix $A$ with dimensions (M$\times$N) has full rank if $\text{rank}(A) = \text{min}(M, N)$.} In the present case, since $\text{rank}(A_{\mathbf{K}})=3$, removing one trial wave function makes $A_{\mathbf{k}}$ a 4 $\times$ 3 matrix with full rank even at $\mathbf{K}$, removing the singularity. However, in those regions of the BZ where the original $A_{\mathbf{k}}$ had full rank, $\text{rank}(A_{\mathbf{K}})=4$, the projection is no longer unitary but semi-unitary, projecting the original target bands onto a lower-dimensional subspace, as discussed in \sref{proj}. Nonetheless, we can still construct a subset of Wannier functions spanning this subspace; we refer to these as “reduced Wannier functions.”

\newcommand\Tstrut{\rule{0pt}{2.9ex}}  % "top" strut
\newcommand\Bstrut{\rule[-1.2ex]{0pt}{0pt}} % "bottom" strut
\begin{table}[t!]
    \centering
    \begin{ruledtabular}
    \begin{tabular}{lccc}
    Minimization Step & $\Omega$ & $\Omega_\textrm{I}$ &  $\widetilde{\Omega}$  \\
     \hline \Tstrut
     P & 0.265 & 0.229 & 0.036 \\
     P+ML & 0.264 & 0.229 & 0.035  \\
     P+SS+P & 0.202 & 0.190 & 0.012 \\
     P+SS+P+ML & 0.201 & 0.190 & 0.011
    \end{tabular}
    \end{ruledtabular}
    \caption{Averaged spreads after each step of the minimization process for the three reduced Wannier functions of the $C=1$ occupied bands for the Haldane model on a $2\times 2$ supercell. The acronyms stand for projection (P), subspace selection (SS), and maximal localization (ML).} 
    \label{tab:spreads}
\end{table}

\begin{figure}[b!]
\begin{center}
\includegraphics[width=3.4in]{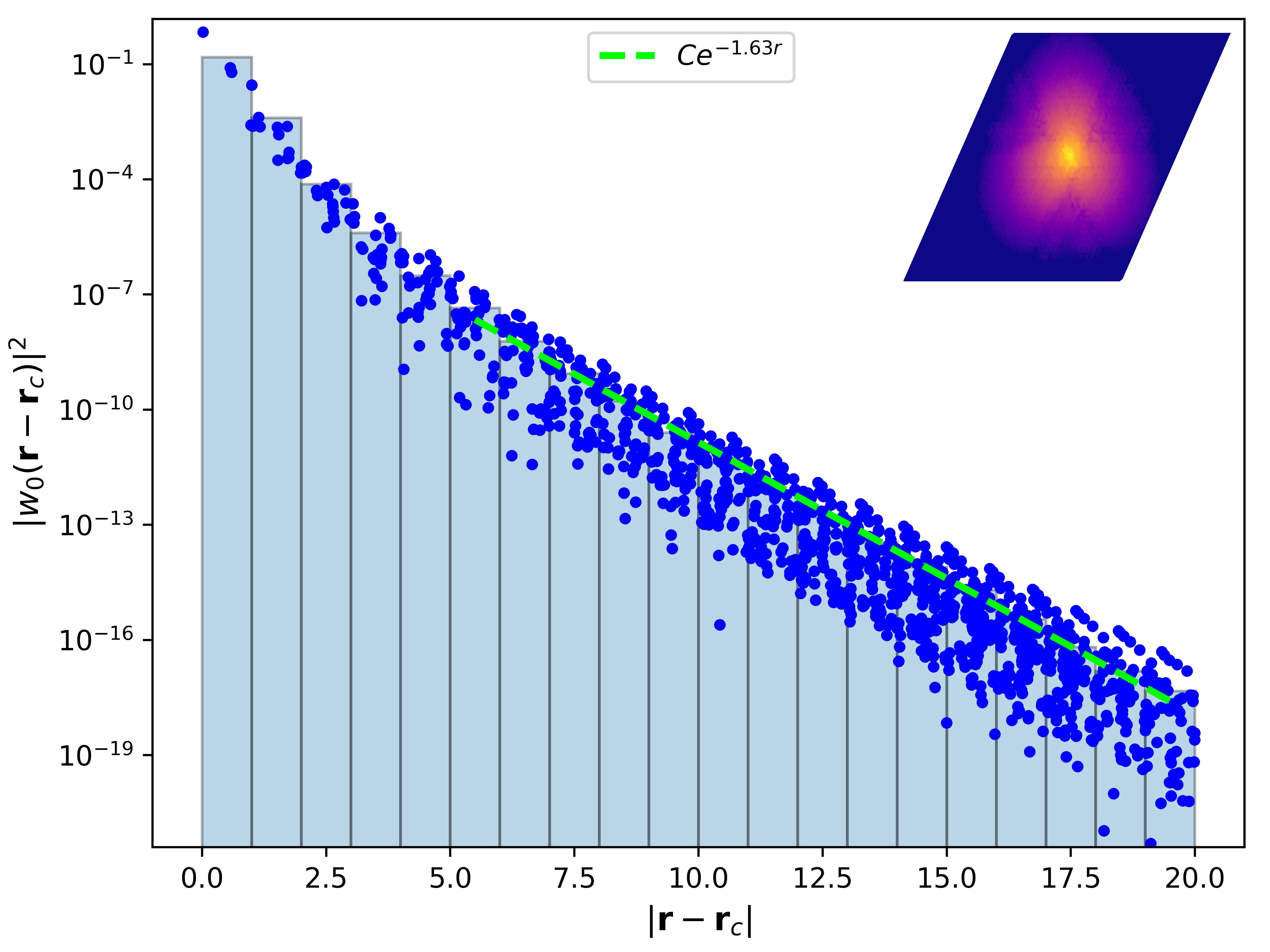}
\end{center}
% \centering
\vspace{-5mm}
\caption{Decay of the electron density (weights $w_0$) as a function of distance from the center of one of the three reduced Wannier functions. These Wannier functions were constructed by projecting onto the three trial functions of \eq{trial} in a 2$\times 2$ supercell, followed by subspace selection and maximal localization. The blue bars are the bin-averaged values of the weights. The inset shows the Wannier density plotted on a log scale in the supercell conjugate to the discrete $k$-mesh.}
\label{FIG3}
% \vspace{-1mm}
\end{figure}

Specifically, we choose three trial wave functions to project onto the four topological bands in Fig.~\ref{FIG1}(b). These trial wave functions are represented by delta functions centered on three of the low-energy sites $\boldsymbol{\tau}_{A,n}$ in the $2\times2$ supercell,
\begin{equation}
     g_n(\mathbf{r}) = \delta(\mathbf{r} - \boldsymbol{\tau}_{A,n}).
     \eqlab{trial}
\end{equation}
Figure~\ref{FIG2}(b) shows the resulting singular values from this projection, revealing that the singularity at the $\mathbf{K}$ point that was present in Fig.~\ref{FIG2}(a) has been eliminated. The smoothness of the singular values suggests that the resulting trial wave functions $\ket{\tilde{\psi}_{n\mathbf{k}}}$ generated from \eq{VWJBlochlike} should be suitable for constructing well-localized Wannier functions.

Having obtained a trivial subspace spanned by these Bloch-like states, we use the methods of Sec.~\ref{sec:MLWF} to minimize the spread of the reduced Wannier functions within this subspace. After projection, we use subspace selection, followed by projection onto the same set of trial wave functions, and finally, maximal localization. This procedure yields three exponentially localized Wannier functions per $2\times2$ cell. Table~\ref{tab:spreads} presents their averaged spreads after each step of the spread minimization, demonstrating the progressive minimization of the gauge-invariant and gauge-dependent spreads. 

\begin{figure}[t!]
\begin{center}
\includegraphics[width=3.4in]{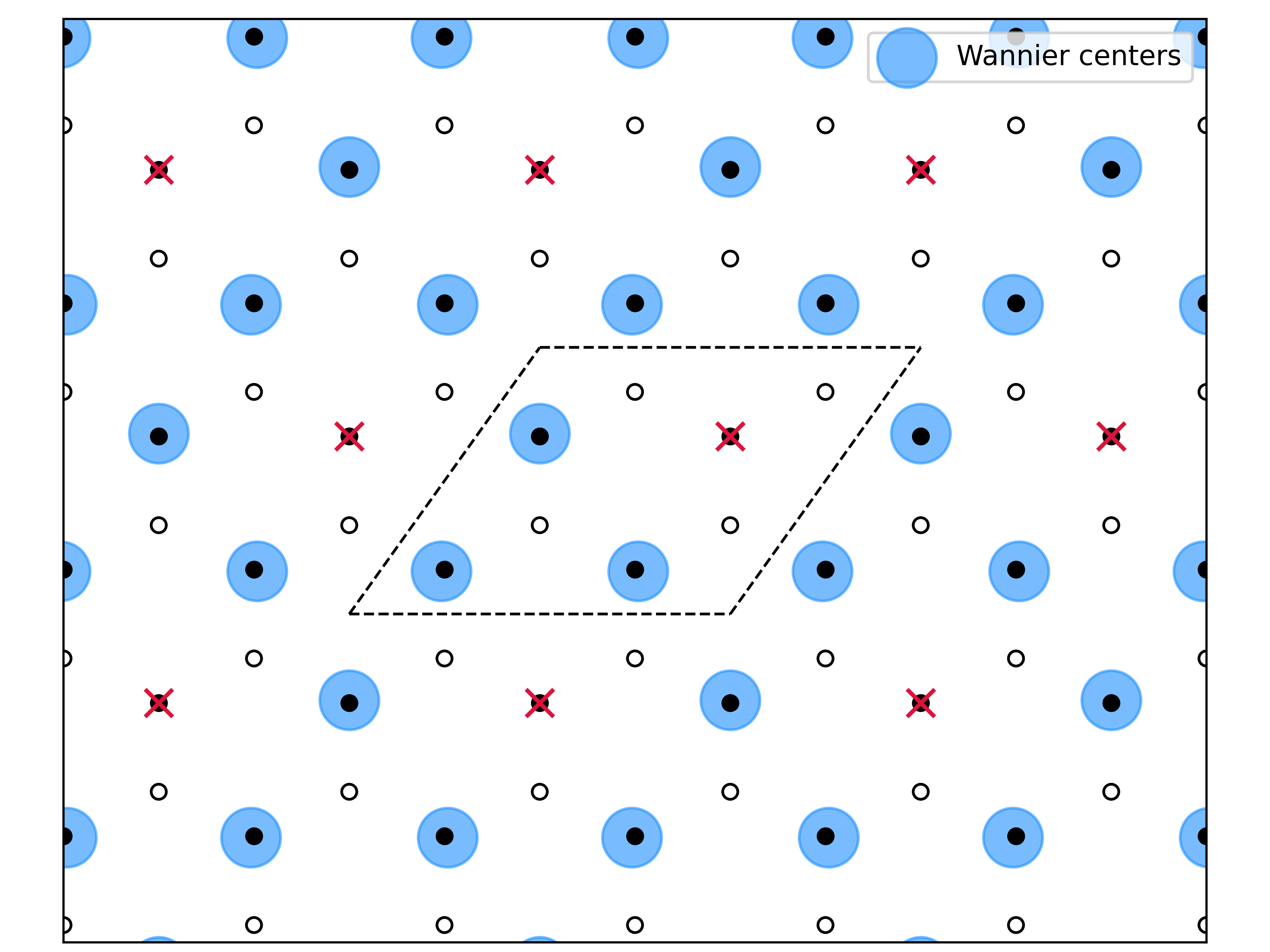}
\end{center}
\vspace{-5mm}
\caption{The centers of the three reduced Wannier functions described in Sec.~\ref{ss:dfcntWF}, indicated by blue circles. The high-energy and low-energy lattice sites are represented by black open and filled circles, respectively. The red ``x"s mark the low-energy site excluded from the set of trial wave functions in \eq{trial}. Dotted lines delineate the supercell.}
\label{FIG4}
\end{figure}

Figure~\ref{FIG3} shows the exponential decay of the density of one of the three Wannier functions away from its center. The other two reduced Wannier functions are related by $C_3$ rotations and show identical exponential decay and spreads. The centers of the three reduced Wannier functions, plotted in Fig.~\ref{FIG4}, are localized on the same low-energy sites as the trial wave functions. Because a trial function has been omitted from the projection, these Wannier functions do not respect the primitive translational symmetry of the lattice. This is a manifestation of the obstruction since, in the topological phase, we cannot generate exponentially localized Wannier functions that respect the lattice symmetries and the symmetries protecting the topology.

\subsection{Disentangled Topological Subspace}

\begin{figure}[t!]
\begin{center}
\includegraphics[width=3.4in]{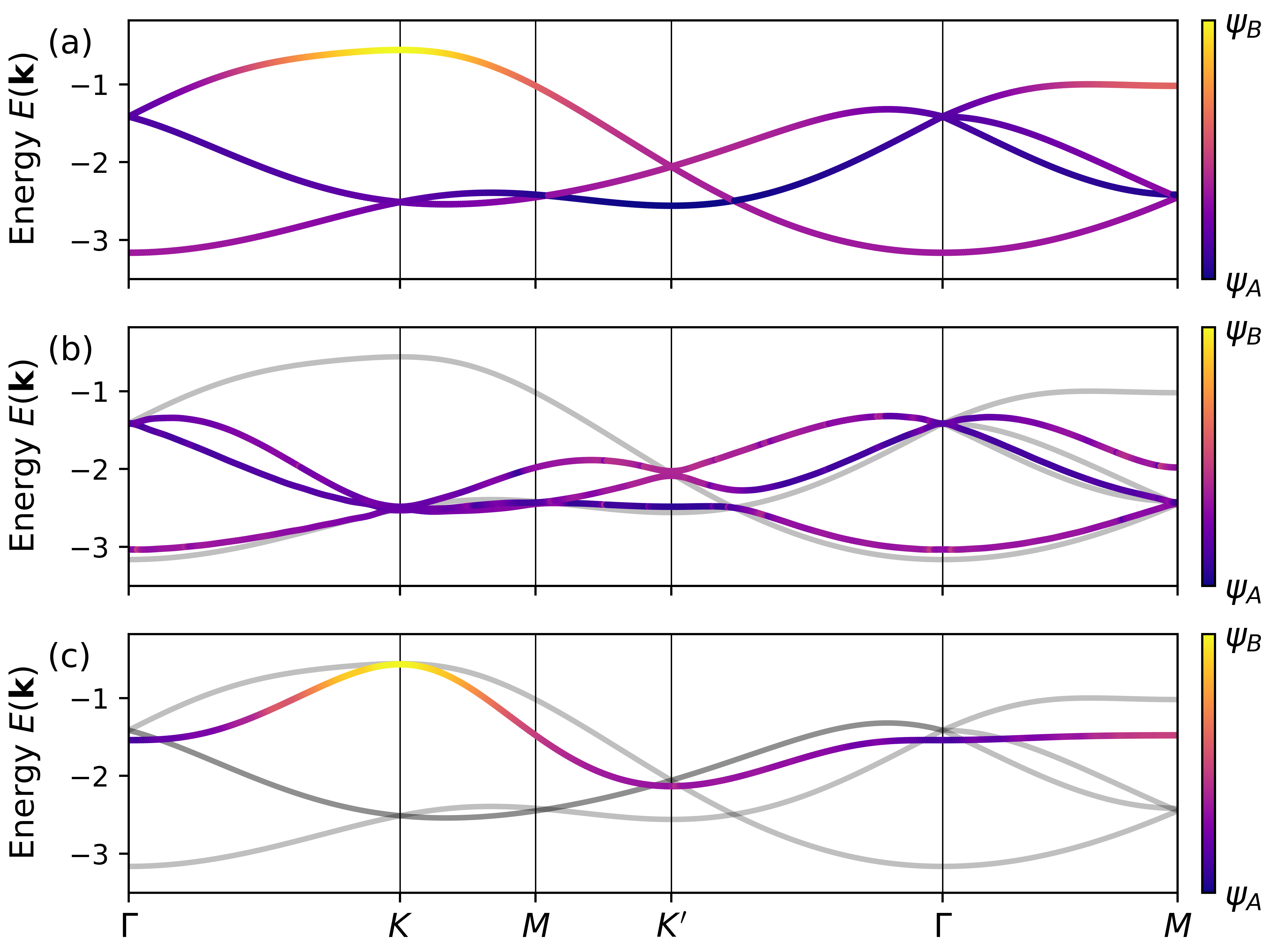}
\end{center}
\vspace{-5mm}
\caption{(a) Folded bands of the topological occupied manifold for the Haldane model on a $2\times 2$ supercell. (b) Interpolated bands of the trivial subspace spanned by the reduced Wannier functions. (c) Interpolated band of the topological subspace complementary to the reduced Wannier functions. As in Fig.~\ref{FIG1}, the color bars correspond to the projection onto low- and high-energy sublattices (purple and yellow respectively).}
\label{FIG5}
\end{figure}

In our construction of reduced Wannier functions, we partition the topological manifold into a subspace free from the obstruction spanned by the reduced Wannier functions and a complementary subspace that inherits the topological character of the original manifold. Symbolically, we have performed the operation
\begin{equation}
    \text{topological} = \text{trivial} \oplus \text{topological},
\end{equation}
through careful use of the projection procedure. The trivial subspace is represented by the projector onto the reduced Wannier manifold $\mathcal{R}$, expressed in terms of the Bloch-like states as, 
\begin{equation}
\eqlab{eq:trivss}
    P^{\text{triv}}_{\mathbf{k}} = \sum_{n=1}^{\dim({\mathcal{R}})} |\tilde{\psi}_{n,\mathbf{k}} \rangle \langle \tilde{\psi}_{n,\mathbf{k}}|.
\end{equation}
These are the states obtained after the initial projection onto trial wave functions, followed by subspace selection and maximal localization. The complement of $P^{\text{triv}}$ in the obstructed manifold is topological, obtained by taking the difference between the occupied and the trivial band projectors, 
\begin{equation}
\eqlab{topss}
P^{\text{topo}}_{\mathbf{k}}= P^{\text{occ}}_{\mathbf{k}} - P^{\text{triv}}_{\mathbf{k}}.
\end{equation}
This subspace retains the topological character of the occupied bands and inherits the obstruction. That is, we have segregated the obstruction into a one-dimensional subspace, leaving the rest of the occupied space free of any topological obstruction.

\def\ssp{{\cal M}} 
For each projector $P_{\mathbf{k}}^\ssp$ onto band manifold $\ssp$ = `occ', `triv', or `topo' in \eq{topss}, we form the projected Bloch Hamiltonian
\begin{equation}
H^\ssp_{\mathbf{k}} = P^\ssp_{\mathbf{k}} H_{\mathbf{k}} P^\ssp_{\mathbf{k}}.
\end{equation}
In the present example, $H_{\mathbf{k}}$ is the Bloch Hamiltonian for the Haldane model on a $2\times 2$ supercell. By diagonalizing the Hamiltonian projected onto each subspace, we obtain the three band structures in Fig.~\ref{FIG5}. The original occupied bands are shown in Fig.~\ref{FIG5}(a), and Fig.~\ref{FIG5}(b)-(c) show the band structures for the three reduced Wannier functions and the single topological band spanning the complementary subspace, respectively. As the trivial and topological subspaces no longer fully contain any particular energy eigenstate at a generic $k$-point, we should not expect the interpolated bands to overlap exactly with the original energy bands. The bands formed by the reduced Wannier functions have a large degree of overlap with the lower-energy bands, and avoid the high-energy band that underwent inversion at $\mathbf{K}$. The band from the topological subspace overlaps strongly with the highest energy-occupied band at the inversion point and shows the same change in character, demonstrating that this state has inherited the topological band inversion of the original energy eigenstates.

It is of interest to inspect the distribution of the Berry curvature and quantum metric in the Brillouin zone for each subspace $P_{\mathbf{k}}^\ssp$. The Berry curvature $\mathcal{B}^{(n)}_{\mu\nu}(\mathbf{k})$ and quantum metric $g^{(n)}_{\mu\nu}(\mathbf{k})$ of band $n\!\in\!\ssp$ are determined respectively by the real and imaginary parts of the quantum geometric tensor \cite{marzari1997, study1905, PATI1991105}
\begin{equation}
    F_{\mu \nu}^{(n)} = \langle \check{\partial}_{\mu} u_{n, \mathbf{k}}| \check{\partial}_{\nu} u_{n, \mathbf{k}} \rangle ,
\end{equation}
where $\check{\partial}_{\mu} = (1-P_{\mathbf{k}}^\ssp)\partial_{\mu}$ is the gauge-covariant derivative and $\partial_{\mu} = \partial_{k_{\mu}}$. Specifically,
\begin{equation}
    \mathcal{B}_{\mu\nu}^{(n)} = -2 \text{Im}F_{\mu \nu}^{(n)}, \quad
    g_{\mu\nu}^{(n)} = \text{Re}F_{\mu\nu}^{(n)} .
\end{equation}

To simplify the presentation of the results, we focus on the band-traced curvature $\mathcal{B}(\textbf{k})=\sum_n \mathcal{B}_{xy}^{(n)}(\textbf{k})$ and the band- and Cartesian-traced metric $\omega_\textrm{I}(\textbf{k})=\sum_n [g_{xx}^{(n)}(\textbf{k})+g_{yy}^{(n)}(\textbf{k})]$, where the band sum runs over $n\!\in\!\ssp$. The notation $\omega_\textrm{I}(\textbf{k})$ is chosen to emphasize the connection to the gauge-invariant spread of the Wannier functions in \eq{OmegaI} through \cite{marzari1997}
\begin{equation}
\Omega_\textrm{I}=\frac{1}{A_{\rm BZ}}\,\int d\textbf{k}\, \omega_\textrm{I}(\textbf{k}),
\end{equation}
where $A_{\rm BZ}$ is the Brillouin zone area. 

\begin{figure}[t!]
\begin{center}
\includegraphics[width=3.4in]{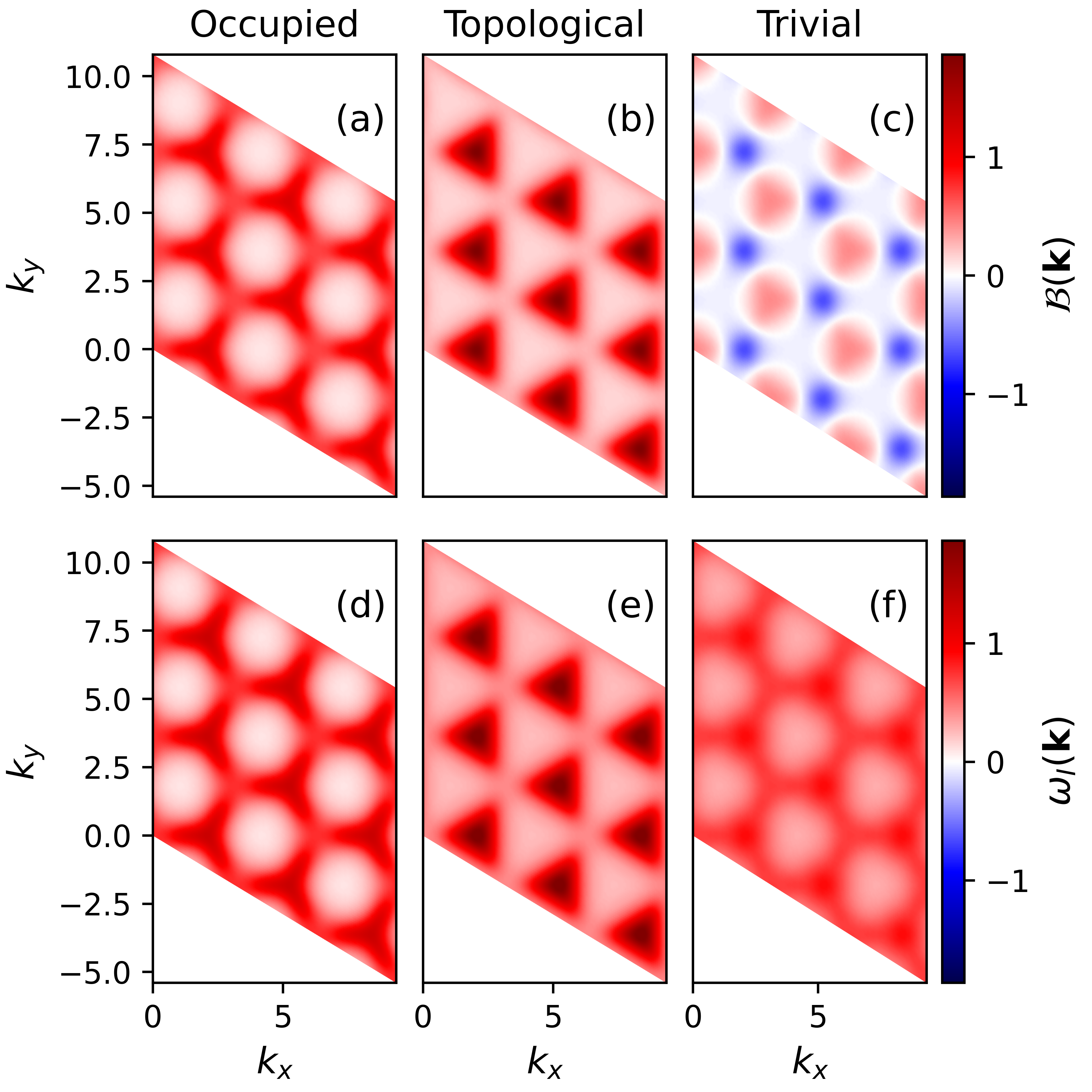}
\end{center}
\vspace{-8mm}
\caption{(a)-(c) Berry curvature $\mathcal{B}(\textbf{k})$ of (a) the occupied subspace, (b) the topological subspace, and (c) the trivial reduced Wannier subspace. (d)-(f) Trace of the quantum metric $\omega_\textrm{I}(\textbf{k})$ of the same subspaces listed above. All plots extend over several reciprocal lattice unit cells.}
\label{FIG6}
\end{figure}

\begin{figure}[t!]
\begin{center}
\includegraphics[width=3.4in]{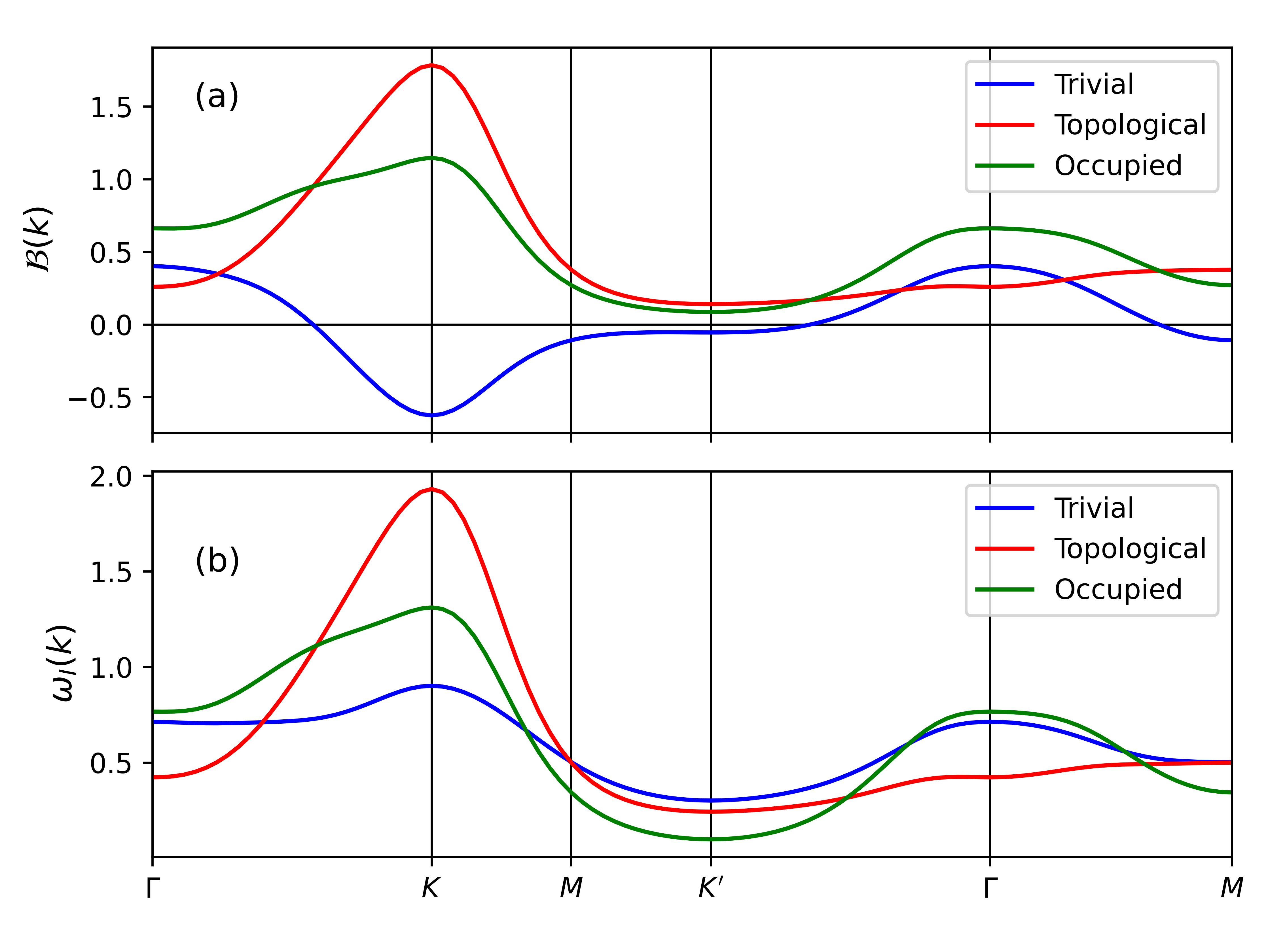}
\end{center}
\vspace{-10mm}
\caption{The interpolation of (a) the Berry curvature and (b) the trace of the quantum metric along a path through high-symmetry points in the BZ. }
\label{FIG7}
\end{figure}

We have computed the Berry curvature $\mathcal{B}(\textbf{k})$ and metric $\omega_\textrm{I}(\textbf{k})$ in the Brillouin zone for the Haldane $2\times2$ supercell in the topological phase. These are plotted for the entire occupied four-band subspace, the topologically obstructed one-band subspace, and the trivial three-band subspace in Figs.~\ref{FIG6} and \ref{FIG7}. For all subspaces, the metric is largest in the vicinity of the band inversion point, where the character of the bands is changing most rapidly. However, it remains finite throughout the BZ in the topological as well as the trivial phase, consistent with Ref.~\cite{thronhauser2006}, which found that the gauge-invariant spread diverges at the topological phase transition but returns to being finite in the topological phase. Upon integrating the Berry curvatures, we confirm that the topological subspace carries a nontrivial Chern number $C=1$, inheriting the Chern number of the occupied subspace, while the trivial subspace has $C=0$. This must be the case since we have obtained exponentially localized Wannier functions for the reduced subspace.

\subsection{Wannier Fraction}
\label{ss:WF}

The number of Wannier functions that can be constructed for a set of trivial bands is simply the number of bands, $N_b$. In the presence of an obstruction, however, the dimension of the subspace that can be spanned by exponentially localized Wannier functions is constrained by the maximum dimension of the null space of $A_{\mathbf{k}}$ at the (possibly multiple) singularities. The null space of $A_{\mathbf{k}}$, denoted $\mathcal{N}(A_{\mathbf{k}})$, contains the states mapped to zero by the projection onto the trial wavefunctions. Its maximum dimension across the BZ, $\mathcal{N}_{\rm max} \equiv \max_{\mathbf{k}}\left[ \text{dim}(\mathcal{N}(A_\mathbf{k})) \right]$, provides the bound on the dimension of the reduced Wannier subspace $\mathcal{R}$,
\begin{equation}
\label{eq: boundR}
    \dim({\mathcal{R}}) \leq N_b - \mathcal{N}_{\rm max}.
\end{equation}
We define the ``Wannier fraction" as the ratio of the number of Wannier functions to the total number of bands in the topological manifold,
\begin{equation}
    f_{W} = \frac{\text{dim}(\mathcal{R})}{N_b},
\end{equation}
quantifying the extent to which the reduced Wannier functions span the topological manifold. This definition combined with Eq.~\ref{eq: boundR} leads to a bound on the Wannier fraction
\begin{equation}
    f_W \leq 1 - \frac{ \mathcal{N}_{\rm max}}{N_b}.
\end{equation}

For a two-band model, we cannot construct Wannier functions for the single occupied band in the topological phase, since $\mathcal{N}_{\rm max} = 1$ leads to $f_W =0$. As we have done above, we must construct a supercell, breaking the primitive translational symmetry, in order to construct a trivial subspace admitting a Wannier representation. When we construct an $N_{\rm sc}\times N_{\rm sc}$ supercell, the dimension of the occupied manifold increases to $n_{\rm occ} = N_{\rm sc}^2$. The dimensionality of the null space at the singularities remains invariant through band folding, and as such, the bound on the Wannier fraction for our two-band model is
\begin{equation}
   f_W \leq 1 - \frac{1}{N_{\rm sc}^2}.
\end{equation}
By constructing larger supercells, we increase the fraction of the topological subspace spanned by the reduced Wannier functions. Our first example presented in the preceding subsections was for the case of a 2$\times$ 2 supercell and $f_W$=3/4. By applying the same procedure to larger supercells, we can construct reduced Wannier functions that span an increasing portion of the topological manifold.

We have demonstrated this for the Haldane model on a $5\times 5$ supercell, which after band folding has $n_{\rm occ} =25$. In the topological phase we still have $C=1$ and $\mathcal{N}_{\rm max} = 1$, giving us a bound of $f_w \leq 24/25$. We compute up to the maximum number of $24$ Wannier functions and examine how the spread of the reduced Wannier functions varies with the dimension of the subspace they span. For each Wannier fraction, we pick trial wave functions localized on randomly chosen subsets of low-energy sites in the supercell. 

\begin{figure}[t]
\begin{center}
\includegraphics[width=3.4in]{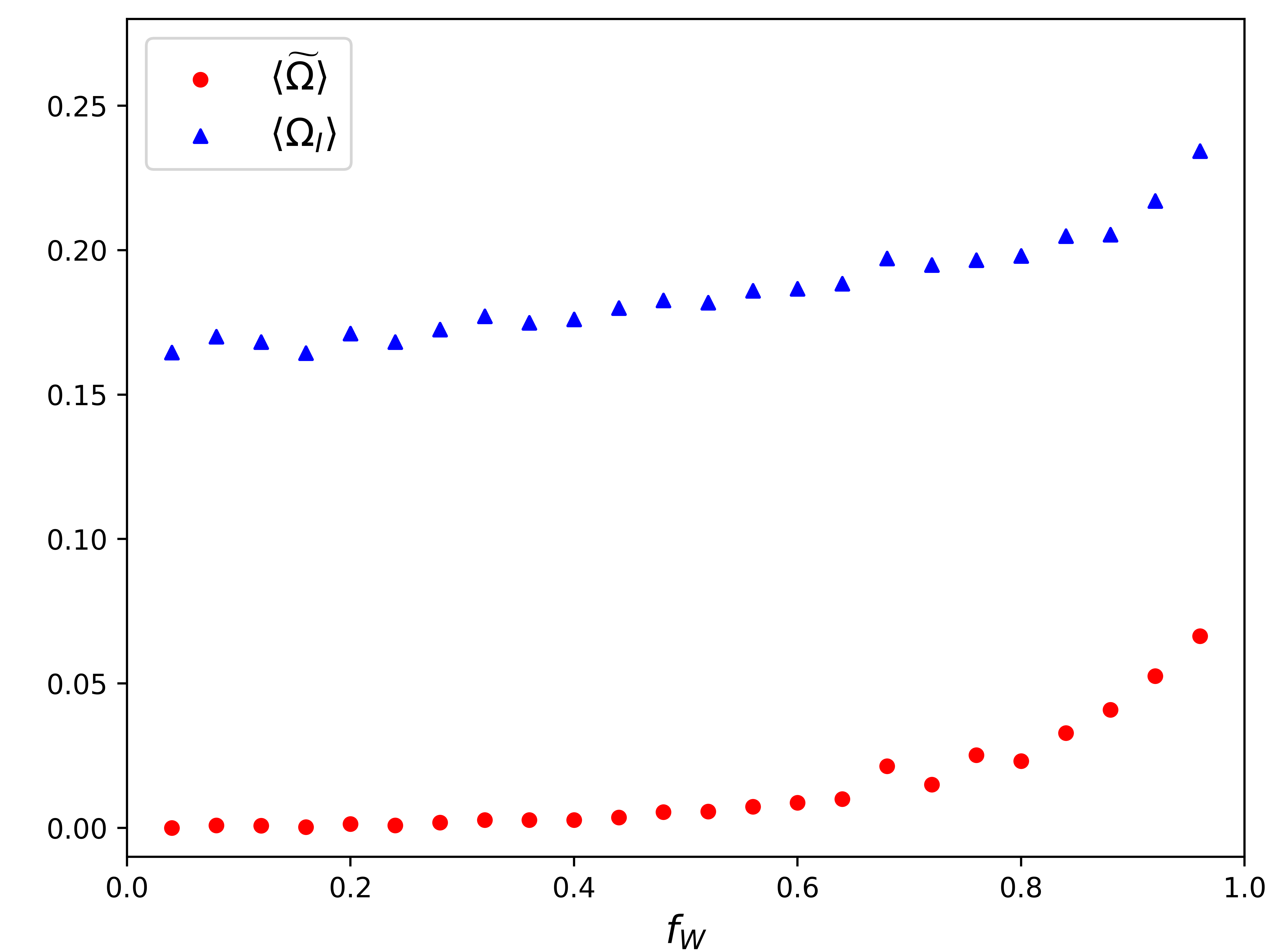}
\end{center}
\vspace{-5mm}
\caption{The dependence of the gauge-invariant (blue triangles) and gauge-dependent (red circles) average Wannier spreads on the Wannier fraction corresponding to the Haldane model for a $5\times 5$ supercell in its $C=1$ phase. The plotted spreads are averaged over the Wannier functions.}
\label{FIG8}
\end{figure}

Fig.~\ref{FIG8} displays the averaged spreads of the reduced Wannier functions for varying Wannier fractions. Notably, the gauge-dependent part of the spread diminishes significantly at low Wannier fractions, while the gauge-independent spread reaches a finite value. As the Wannier fraction approaches saturation ($f_W=1$), the spreads increase rapidly, highlighting the trade-off between the reduced Wannier functions spanning larger fractions of the topological manifold and their localization.

\begin{figure}[t]
\begin{center}
\includegraphics[width=3.4in]{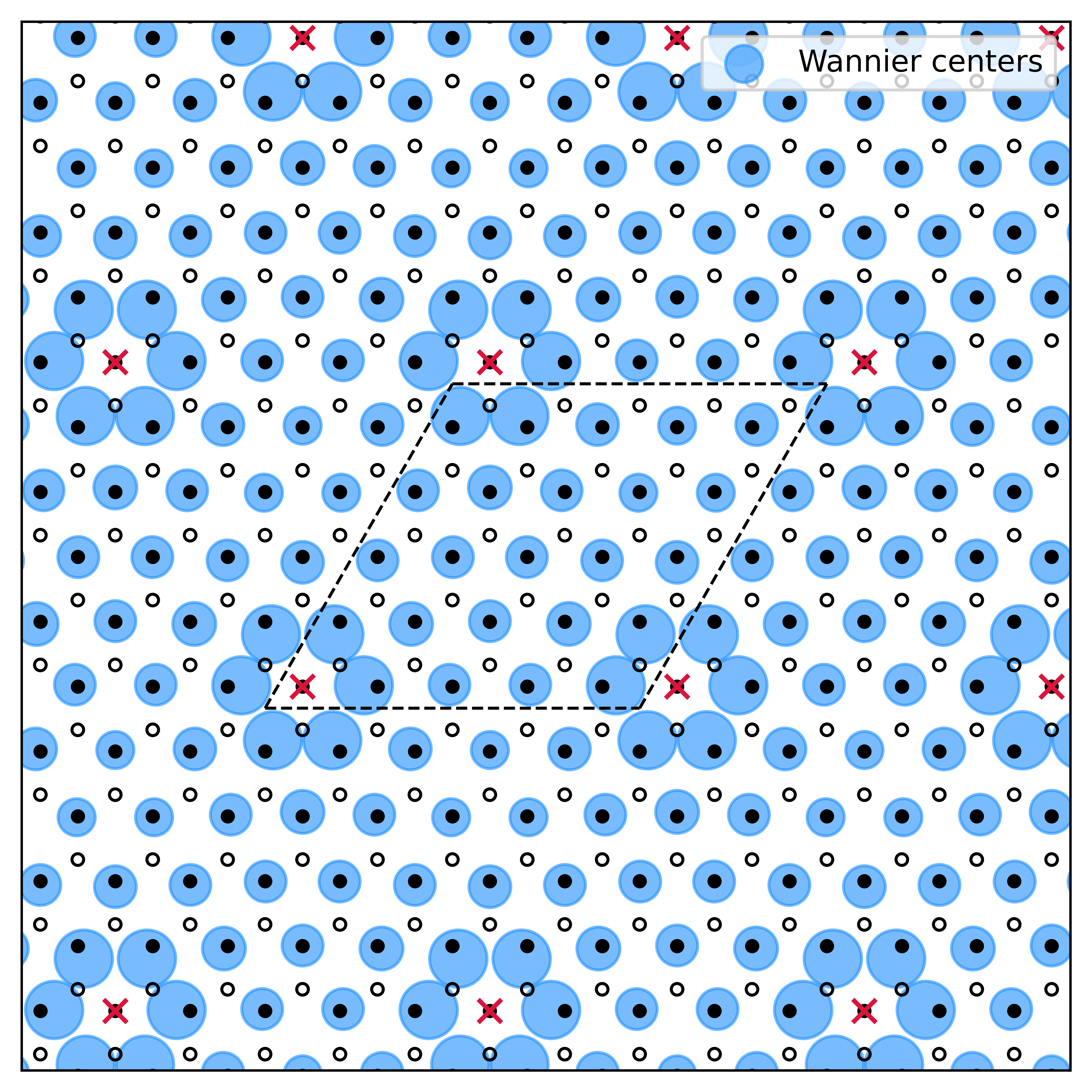}
\end{center}
\vspace{-5mm}
\caption{Wannier functions (blue circles) at Wannier fraction $f_W = 24/25$. Red ``x"s denote the sole low-energy site on which there is no trial wave function. The sizes of the blue circles reflect the Wannier spreads.}
\label{FIG9}
\end{figure}

The Wannier centers corresponding to the largest Wannier fraction computed, $f_W = 24/25$, are depicted in Fig.~\ref{FIG9}. The low-energy site omitted from the trial wave functions is a center of $C_3$ symmetry and also approximate $C_6$ symmetry from the perspective of the Wannier centers. The Wannier centers exhibit an apparent attraction towards the omitted site, resulting in a sizeable shift from their nominal locations. This reduction of symmetry is one of the consequences of the topological obstruction. The spreads of the centers closest to the omitted site are larger than those of the other centers. Heuristically, these observations are consistent with a picture in which the neighboring Wannier functions are uniformly ``filling in" for the omitted one in order to minimize the total spread.

\section{Reduced Wannier Functions for $\mathbb{Z}_2$ Insulators}
\label{redwf_Z2}

In time-reversal invariant 2D and 3D insulators with a nontrivial ($\mathbb{Z}_2$-odd) index, there is also a topological obstruction that prevents the existence of a smooth gauge that respects time-reversal symmetry and thus forbids the construction of Wannier functions that all take the form of Kramers pairs \cite{soluyanov2011}. Here, we can use a procedure analogous to that of the preceding subsection, where we decompose the 2$N$-dimensional occupied space into a $2(N\!-\!1)$-dimensional trivial manifold represented by Kramers-pair Wannier functions, relegating one topological pair of bands to the remaining subspace.

\subsection{Kane-Mele Model}

\begin{figure}[t!]
\begin{center}
\includegraphics[width=3.4in]{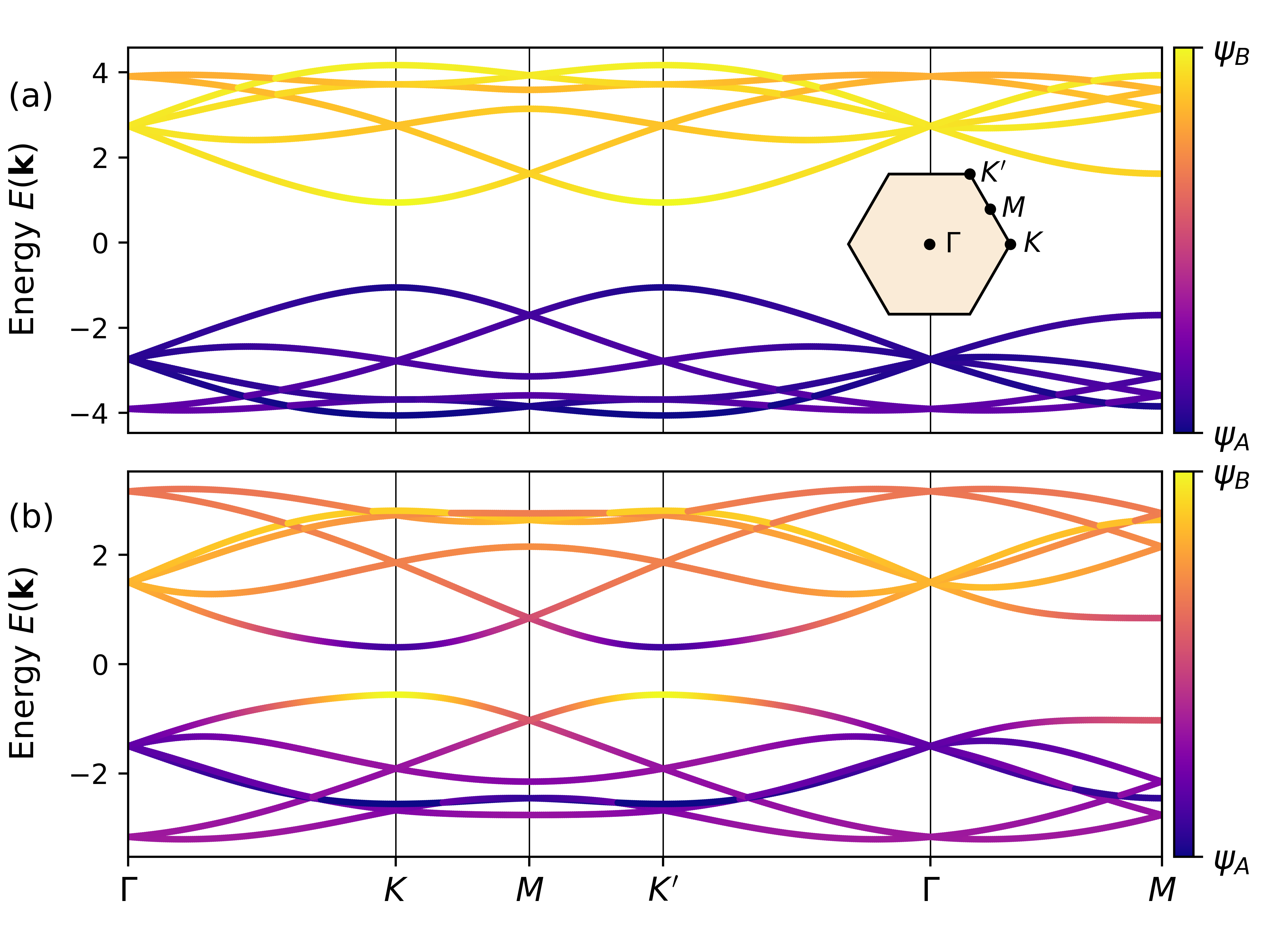}
\end{center}
% \centering
\vspace{-10mm}
\caption{Bands for the Kane-Mele model on a $2\times 2$ supercell. The color indicates the projection of the eigenstates onto sublattices, with purple and yellow representing the low- and high-energy sublattices, respectively. (a) The bands in the trivial phase ($\mathbb{Z}_2=0$) where $\Delta = 2.5$. (b) The bands in the topological phase ($\mathbb{Z}_2=1$) where $\Delta = 1$. In both cases, $t_1 = 1$, $\lambda_{\textrm{SO}} = 0.3t$, $\lambda_{\textrm{R}}=0.25t$.}
\label{FIG10}
% \vspace{-1mm}
\end{figure}

To demonstrate the analogous approach for constructing reduced Wannier functions in the context of $\mathbb{Z}_2$ insulators, we will use the Kane-Mele tight-binding model describing a 2D $\mathbb{Z}_2$-odd insulator for certain parameter regimes. Like the Haldane model, the Kane-Mele model describes a honeycomb lattice with the same primitive lattice vectors and site locations, but now each site hosts a spinor degree of freedom. The tight-binding Hamiltonian takes the form

\begin{equation}
\begin{split}
   H = &\  t_1 \sum_{\langle i, j \rangle} (c_i^{\dagger}c_j + \textrm{h.c.}) +   \lambda_{\textrm{SO}} \sum_{\langle\langle i,j \rangle\rangle} (ic_i^{\dagger} \sigma_z c_j + \textrm{h.c.})
    \\ & + \lambda_{R} \sum_{\langle i,j\rangle} (i c_i^{\dagger} \boldsymbol{\sigma} \times \hat{\mathbf{d}}_{\langle i, j \rangle} c_j + \textrm{h.c.}) + \Delta\sum_i (-)^i  c_i^{\dagger}c_i
\end{split}   
\end{equation}
with staggered onsite energies $\pm \Delta$ defining the high-energy sublattice $B$ and low-energy sublattice $A$, nearest neighbor hopping term $t_1$, second-nearest neighbor spin-orbit interaction $\lambda_{\textrm{SO}}$, and a Rashba term $\lambda_{\textrm{R}}$. The spin labels have been suppressed on the creation and annihilation operators, and the spin indices are contracted over a Pauli matrix if it appears and over the identity if not. The unit vector $\hat{\mathbf{d}}_{\langle i, j \rangle}$ points in the direction from site $j$ to site $i$. A gap closure occurs at the $\mathbf{K}$ and $\mathbf{K'}$ points simultaneously when $\Delta =\pm \sqrt{(3\sqrt{3}\lambda_{\textrm{SO}})^2-4\lambda_{\textrm{R}}^2}$, transitioning the system between $\mathbb{Z}_2$-even and $\mathbb{Z}_2$-odd. Henceforth, we set $t_1=1$, $\lambda_{\textrm{SO}}=0.3$, $\lambda_{\textrm{R}}=0.25$, in which case the $\mathbb{Z}_2$-odd phase occurs when $|\Delta| < 1.47$. We present results for the $\mathbb{Z}_2$-even phase with $\Delta = 2.5$, and the $\mathbb{Z}_2$-odd phase with $\Delta = 1.0$.

\subsection{Reduced Wannier Functions}

\begin{figure}[t!]
\begin{center}
\includegraphics[width=3.4in]{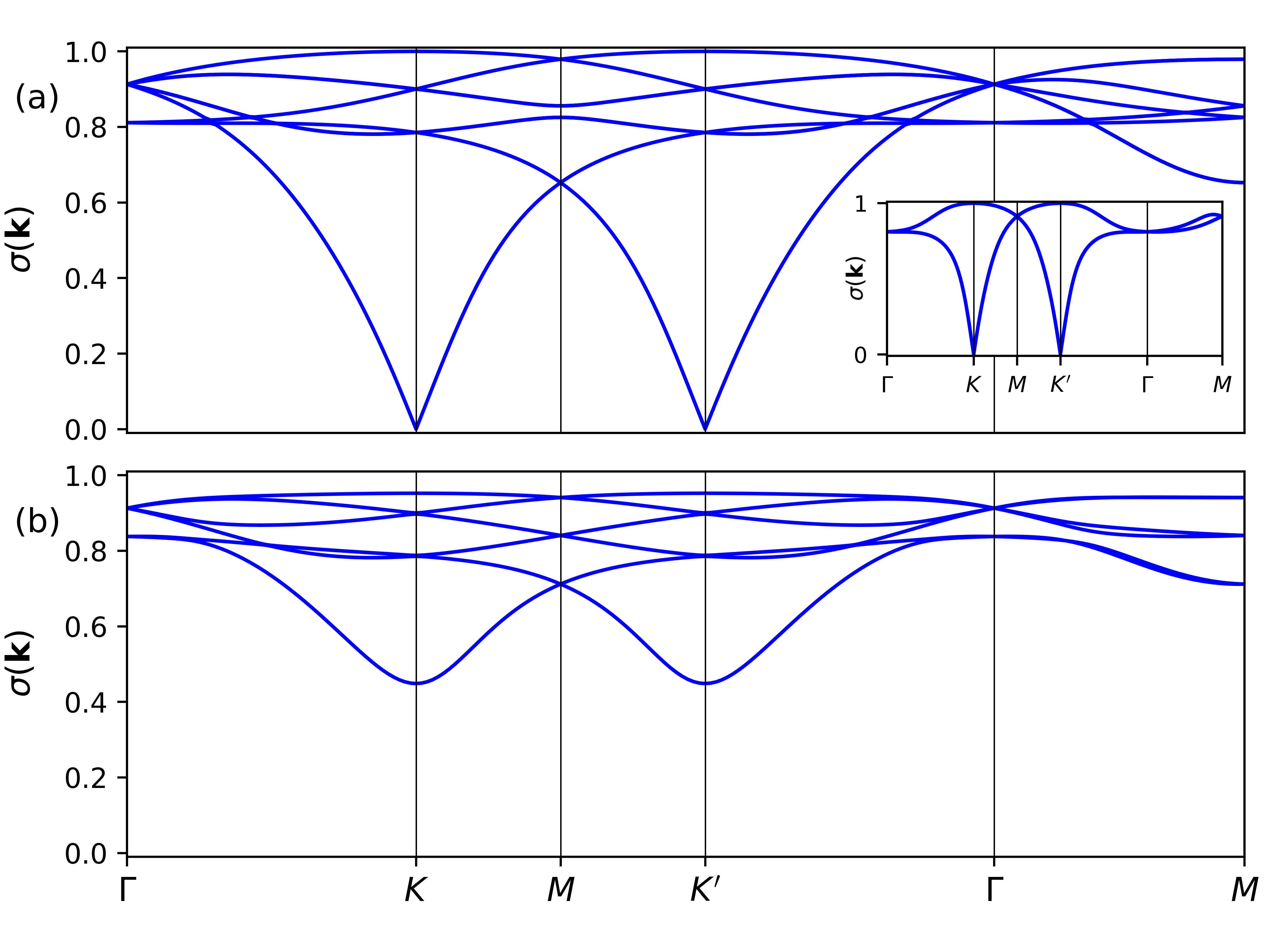}
\end{center}
% \centering
\vspace{-10mm}
\caption{The singular values plotted along a high-symmetry path in the BZ corresponding to the Kane-Mele model in the $\mathbb{Z}_2$-odd phase for a $2\times2$ supercell. (a) All eight trial functions from Eq.~(\ref{EQ:trialkm}) were used in the projection. The inset shows the singular values in the case of a single primitive cell, with the trial wave functions being delta functions on the sole low-energy site spin-polarized along $\pm z$. (b) Six delta functions on a subset of three low-energy sites are used as trial wave functions, all spin-polarized along $\pm z$.} 
\label{FIG11}
% \vspace{-1mm}
\end{figure}

As we did for the Haldane model, we will construct a $2\times 2$ supercell, in which case there are $8$ occupied bands. The band structure is shown in Fig.~\ref{FIG10}, where we see in Fig.~\ref{FIG10}(b) that in the $\mathbb{Z}_2$-odd phase there are band inversions at the $\mathbf{K}$ and $\mathbf{K}'$ points where the gap closed at the topological phase transition. In the trivial phase, we expect the occupied eigenstates to be mostly localized on the low-energy sublattice, as indicated by the sublattice projection color scale in Fig.~\ref{FIG10}(a). As such, we opt for the natural choice of trial wavefunctions localized on the low-energy sublattice, spin-polarized along $\pm z$,
\begin{equation}
\label{EQ:trialkm}
   | g_{\pm,n}\rangle = \delta(\mathbf{r}-\boldsymbol{\tau}_{A,n})\,|\pm z\rangle .
\end{equation}
If we try to use the projection procedure using this full set of trial wavefunctions while in the $\mathbb{Z}_2$-odd phase, we get the same type of divergence as in the case of the Chern insulator, shown here for the $\mathbb{Z}_2$-insulator in Fig.~\ref{FIG11}. This is again owing to the band-inversion at the $\mathbf{K}$ and $\mathbf{K}'$ points, making the highest-energy occupied bands at these points orthogonal to the set of trial wavefunctions. 

\begin{figure}[t!]
\begin{center}
\includegraphics[width=3.4in]{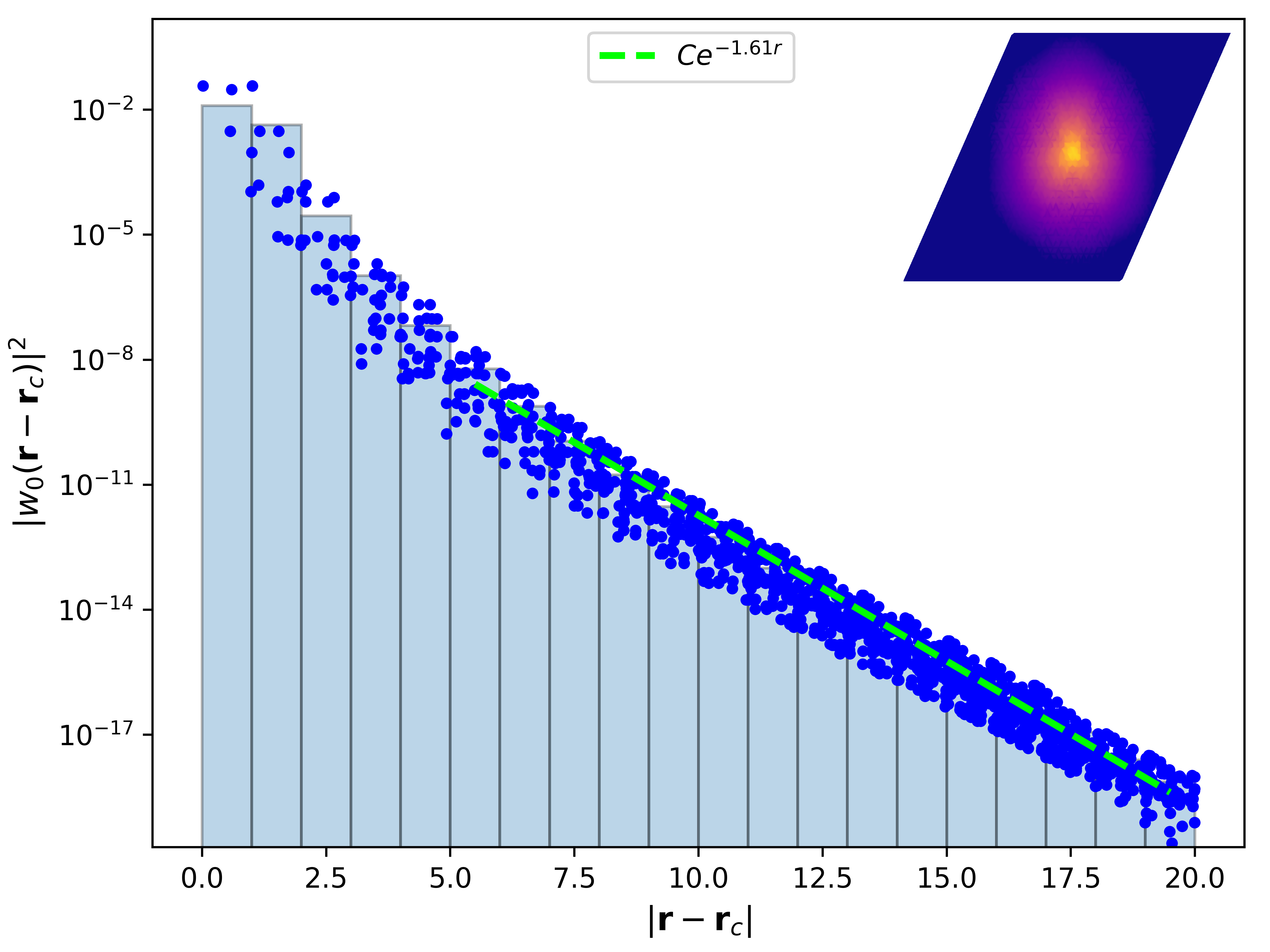}
\end{center}
% \centering
\vspace{-5mm}
\caption{Decay of the electron density (weights $w_0$) as a function of distance from the center of one of the three reduced Wannier functions. These Wannier functions were constructed by projecting onto the six trial functions of Eq.~\ref{EQ:trialkm} in a 2$\times 2$ supercell, followed by subspace selection and maximal localization. The blue bars are the bin-averaged values of the weights. The inset shows the Wannier density plotted on a log scale in the supercell conjugate to the discrete $k$-mesh.}
\label{FIG12}
% \vspace{-1mm}
\end{figure}

Instead, we must now omit the two spin-polarized trial wavefunctions from one of the low-energy sites in the $2\times2$ supercell in order to simultaneously resolve the singularities at both $\mathbf{K}$ and $\mathbf{K'}$. Performing the projection followed by subspace selection and maximal localization gives us three Kramers pairs of exponentially localized Wannier functions. Each of the pairs are localized on one of the low-energy sites populated by trial wavefunctions. The decay of the density of one of the Wannier away from its center is plotted in Fig.~\ref{FIG12}, showing its exponential localization.

\subsection{Disentangled Topological Subspace}

\begin{figure}[t!]
\begin{center}
\includegraphics[width=3.4in]{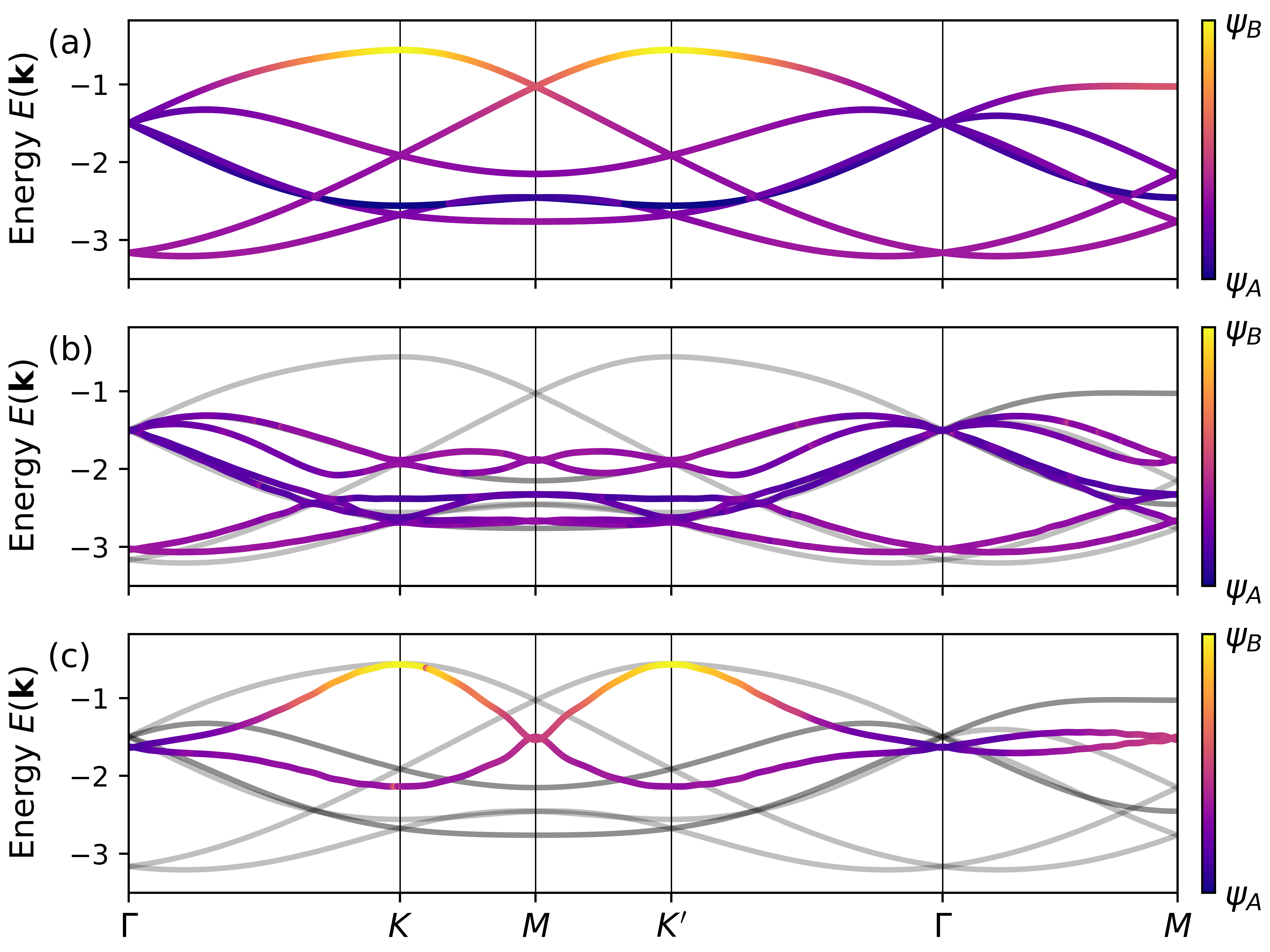}
\end{center}
\vspace{-5mm}
\caption{(a) Folded bands of the topological occupied manifold for the Kane-Mele model on a $2\times 2$ supercell. (b) Wannier interpolated bands of the trivial subspace spanned by the reduced Wannier functions. (c) Interpolated bands of the topological subspace complementary to the reduced Wannier functions. The color bars correspond to the projection onto low- and high-energy sublattices (purple and yellow, respectively).}
\label{FIG13}
\end{figure}

Having obtained the reduced Wannier functions, we have constructed a six-band trivial subspace representable by exponentially localized Wannier functions from an eight-band topologically obstructed occupied subspace of a $\mathbb{Z}_2$-odd insulator. The complementary band projector from \eq{topss} defines the subspace spanned by the two disentangled topological bands from the obstructed occupied manifold. From these band projectors, we construct the Hamiltonian projected onto these subspaces and find the projected bands shown in Fig.~\ref{FIG13}. The trivial bands corresponding to the reduced Wannier functions overlap substantially with the lower energy bands, avoiding the bands that underwent inversion at the topological phase transition. Conversely, the two topological bands show the same inverted character as the highest energy-occupied band and overlap exactly with it at the band inversion point.

\subsection{Wannier Fraction}

The concept of Wannier fractions and supercell construction is naturally extended to the case of $\mathbb{Z}_2$ insulators. We now must omit two trial wavefunctions from the obstructed manifold to resolve the rank deficiency at the singularities, giving a new bound on the Wannier fraction,

\begin{equation}
   f_W \leq 1 - \frac{2\mathcal{N}_{\rm max}}{n_{\rm occ}}.
\end{equation}

\begin{figure}[b!]
\begin{center}
\includegraphics[width=3.4in]{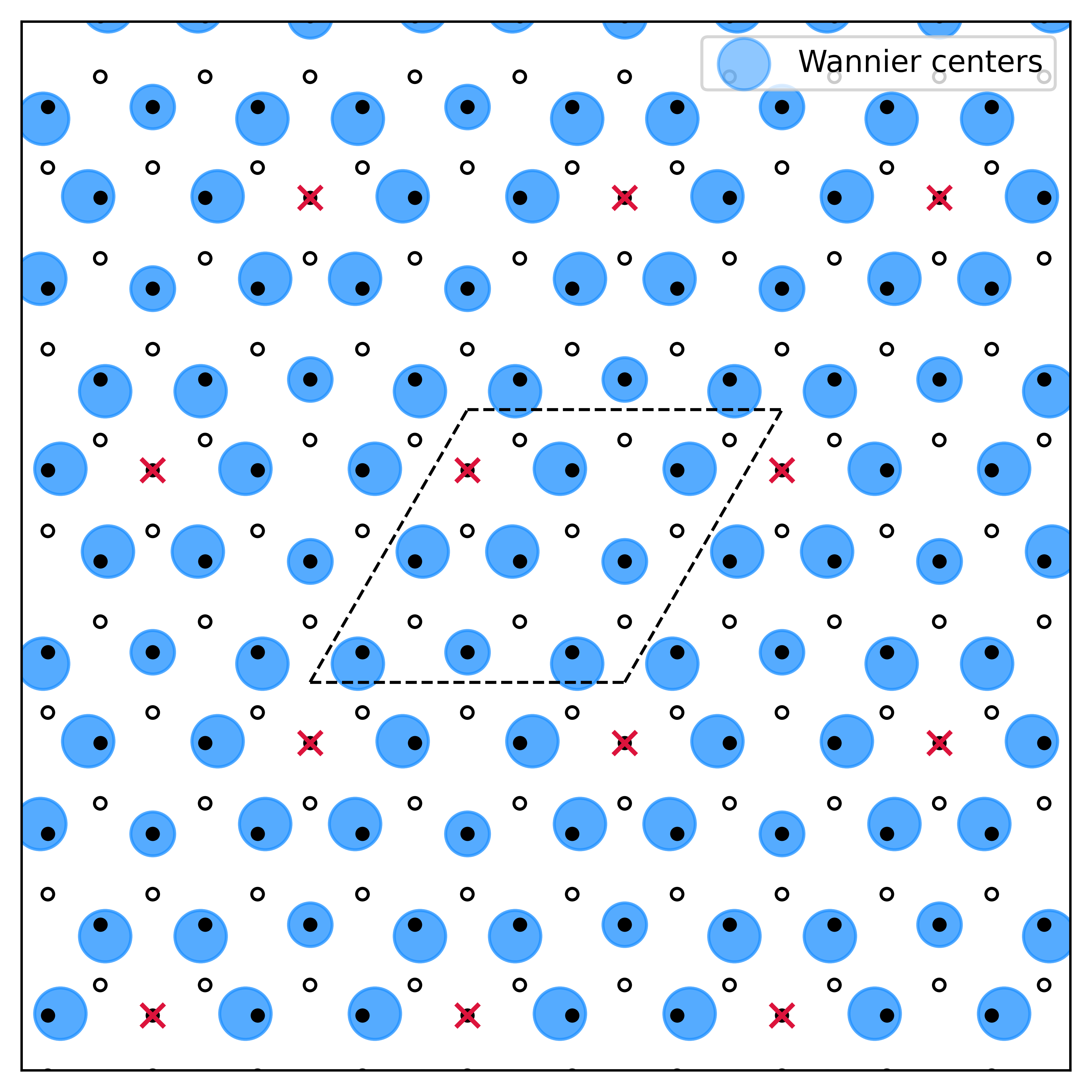}
\end{center}
\vspace{-5mm}
\caption{Wannier functions (blue circles) at Wannier fraction $f_W = 16/18$. Red ``x''s denote the sole low-energy site on which there is no trial wave function. The sizes of the blue circles reflect the Wannier spreads.}
\label{FIG14}
\end{figure}

In the $\mathbb{Z_2}$-odd phase, the dimension of the null-space of the overlap matrix formed by trial wavefunctions and Bloch eigenstates is maximally $\mathcal{N}_{\rm max}=1$, so that $f_W \leq 1 - 2/n_{\rm occ}$. As an example, we have put the Kane-Mele model on a $3\times 3$ supercell, giving us $n_{\rm occ} = 18$, and constructed the reduced Wannier functions for the maximal fraction $f_W = 16/18$ in the $\mathbb{Z}_2$-odd phase. The Wannier functions come in Kramers pairs and have exponentially localized tails. Fig.~\ref{FIG14} shows the resulting Wannier centers. As before, we see that the Wannier centers close to the omission site are shifted away from their nominal locations and otherwise are closely localized on the sites of the trial wavefunctions.
\\
\\
\\
\\
\clearpage

\section{Discussion}

We have demonstrated a procedure for constructing a set of exponentially localized Wannier functions that span a subspace of an obstructed set of topological bands. In our approach, the topological obstruction is isolated—confined to a one-band subspace in a $C=1$ Chern insulator or to a two-band subspace in a $\mathbb{Z}_2$-odd insulator— leaving the remaining part of the obstructed manifold topologically trivial and thus Wannierizable. Through band folding, we can increase the dimensionality of the trivial subspace, increasing the coverage of the reduced Wannier functions over the obstructed manifold at the cost of breaking translational symmetry and decreased localization. Although demonstrated for a half-filled two-band Chern insulator, the method is readily applicable to multiband systems and those with higher Chern number. In the former case, it may not even be necessary to break translational symmetry in order to segregate the obstruction. In the latter case, it would be interesting to test whether the stronger obstruction can still easily be isolated to a single topological band. Similar remarks apply to time-reversal invariant systems with a nontrivial $\mathbb{Z}_2$ index.

A crucial aspect of our approach is the deliberate exclusion of the subspace responsible for the net Chern number from the Wannier basis. Since this topological sector cannot be represented by exponentially localized Wannier functions, we instead capture as much of the remaining manifold with exponentially localized Wannier functions as possible. This contrasts with previous techniques that forced a Wannier representation by incorporating bands of opposite topology. While this can be a valid strategy in many contexts, there are scenarios—such as when a band of opposite topology does not exist within a relevant energy window—where this is not desirable. For instance, if bands around the Fermi level have $C=-1$ and no $C=+1$ manifold lies nearby in energy, then incorporating a distant conduction band would introduce largely unwanted states into the Wannier representation. Other methods have retained the topological contribution in Chern insulators \cite{Gunawardana2024, li2024}, resulting in Wannier functions with polynomial tails. Here, we deliberately leave the topological contribution out of the Wannier description, obtaining exponentially localized Wannier functions. The topological subspace remains fully accessible for incorporation into the complete basis for any operator of interest. One could even view the procedure as its inverse, removing the trivial subspace to isolate the nontrivial one, depending on the requirements of the physical problem at hand.

Our work has been motivated in part by the recent flowering of interest in flat-band systems, often realized in moiré graphene or transition-metal dichalcogenide materials, where exotic correlated phases have been reported, including superconducting, Wigner crystal, quantum anomalous Hall, and fractional quantum anomalous Hall states \cite{Shayegan2022, siddharth2013, yang2024, zhao2012, kato2022, wu2007, Jaworowski2018, sheng2011}. In these systems, the flatness of the bands diminishes the kinetic energy, suppressing the tendency toward the formation of a Fermi liquid relative to more unusual many-body states. Depending on the microscopic details, the relevant correlated state may predominantly occupy either the trivial or topological subspace of the flat-band manifold. With our approach, one could separate the manifold into a topological and trivial part, segregating the topological part for analyzing, e.g., anomalous Hall effects, and the trivial part for the formation of trivial Mott-like behaviors. In any case, having a Wannier representation for the trivial portion of the band manifold can be a valuable tool, potentially simplifying the construction and analysis of strongly correlated states in these flat-band settings.

More specifically, we anticipate that our approach may have an impact in the field of 2D flat-band (e.g., moir\'e) systems at partial filling. While single-particle energies alone would dictate metallic filling of the lowest-energy eigenstates, interaction effects tend to dominate in these systems, driving the selection of an occupied subspace based on other criteria. In some cases this might lead to charge density waves (CDWs) with broken translational symmetry commensurate with the fractional filling, with occupation of either a topological or a trivial subspace of the band space, depending on details. By constructing supercells commensurate with the CDWs, one could possibly remove the obstruction in the same way we have done here for Chern and $\mathbb{Z}_2$-odd insulators. This could lead to new insights into the formation of states exhibiting the fractional quantum Hall effect in such systems. 

There is an inherent bias in choosing a specific set of trial functions, as is typical for Wannierization via projection. In our model, there is no obvious choice of which trial wavefunction to omit, or in other words, where to put the lattice defect(s) where no localized Wannier function exists. However, in other models there may be a particular subset of sites where one expects localized Wannier functions. For example, in twisted bilayer graphene, at the magic angle the charge density exhibits a ``fidget spinner" pattern where it is localized on triangular sites in the moiré unit cell \cite{Kang2018}. In the four-band model, the reduced Wannier functions could be chosen to sit on these three sites, potentially resolving the non-trivial topology in the flat bands. In this case, there is an obvious choice of trial wavefunctions for constructing the reduced Wannier functions.

In systems with fragile obstructions, such as those occurring in some moir\'e systems, obtaining a minimal Wannier representation of the obstructed bands is challenging due to the necessity of including additional trivial bands to trivialize the obstructed subspace \cite{po2019}. Our approach would instead reduce the dimensionality of the band manifold containing the obstruction. It remains to be seen whether our approach, or a derivative of it, could be profitably applied to such systems. In the context of crystalline topological insulators, the corresponding issue is the inability to form Wannier functions that are permuted into one another by the crystallographic symmetry operations; again, it may be possible to do this for most, but not all, of the occupied bands.

Of course, as the Bloch-like states corresponding to the reduced Wannier functions are not energy eigenstates, they cannot be used to compute spectral properties. However, expectation values of one-particle operators, such as energy or charge density, can still be written as a sum of traces over the trivial and topological subspaces, with the potential advantage that the trivial part can be computed in a Wannier basis. Our method of decomposing the obstructed bands into a localized trivial subspace and a delocalized topological subspace is similar in spirit to that employed in Refs.~\cite{song2022} and \cite{yu2023}. In these models of twisted bi- and tri-layer graphene, the (fragile) obstructed flat bands are partitioned into a set of trivial, highly localized heavy fermions that hybridize with the itinerant conduction bands. In these cases, the fragile obstruction is mitigated by incorporating higher-energy trivial bands. However, in many moiré models the flat bands may carry a non-zero Chern number, rendering such a resolution infeasible and highlighting the potential utility of our approach.

Finally, although we have illustrated our approach starting from simplified tight-binding models, it can be extended straightforwardly to first-principles calculations based on density functional theory (DFT). Modern first-principles workflows commonly involve constructing Wannier functions using tools like \textsc{Wannier90} \cite{MOSTOFI20142309, Marrazzo2023TheWS}. The projection and disentanglement procedure we describe are already integral parts of such workflows. As a result, it should be possible to isolate the topological obstruction and Wannierize the remaining subspace in a first-principles context, just as was done here for tight-binding models.

 \section*{Acknowledgments}
%=================================================

This work was supported by NSF Grant DMR-2421895. We thank Martin Claassen for useful discussions that stimulated our interest in this project.

\newpage
\bibliography{ref}

\end{document}